\documentclass[aps,onecolumn,groupedaddress,nofootinbib,showpacs]{revtex4}

\usepackage{graphicx}
\usepackage{bm}
\usepackage{listings}
\usepackage{url}

\begin{document}


\title{Molecular Simulation of Electrode-Solution Interfaces}
\date{Aug. 2020}

\author{Laura Scalfi,$^1$ Mathieu Salanne,$^{1,2}$ and Benjamin
Rotenberg$^{1,2}$}

\affiliation{$^1$Sorbonne Universit\'e, CNRS, Physicochimie des \'electrolytes
et Nanosyst\`emes Interfaciaux, F-75005 Paris, France; email:
benjamin.rotenberg@sorbonne-universite.fr \\
$^2$R\'eseau sur le Stockage Electrochimique de l'Energie (RS2E), FR CNRS 3459, 80039 Amiens Cedex, France
}


\begin{abstract}
Many key industrial processes, from electricity production, conversion and storage to electrocatalysis or electrochemistry in general, rely on physical mechanisms occurring at the interface between a metallic electrode and an electrolyte solution, summarized by the concept of electric double layer, with the accumulation/depletion of electrons on the metal side and of ions on the liquid side. While electrostatic interactions play an essential role on the structure, thermodynamics, dynamics and reactivity of electrode-electrolyte interfaces, these properties also crucially depend on the nature of the ions and solvent, as well as that of the metal itself. Such interfaces pose many challenges for modeling, because they are a place where Quantum Chemistry meets Statistical Physics. In the present review, we explore the recent advances on the description and understanding of electrode-electrolyte interfaces with classical molecular simulations, with a focus on planar interfaces and solvent-based liquids, from pure solvent to water-in-salt-electrolytes.
\end{abstract}


\maketitle


\tableofcontents

\section{INTRODUCTION}

Many key industrial processes, from electricity production, conversion and storage~\cite{salanne2016a}, to electrocatalysis or electrochemistry in general~\cite{seh_combining_2017}, rely on physical mechanisms occurring at the interface between a metallic solid (electrode), allowing the transport of electrons, and an electrolyte solution, in which electric currents may arise from the transport of ionic species or the orientation of polar molecules. The most interesting features of such electrode-solution interfaces emerge from the coupling between the charges accumulated on both sides, summarized by the concept of ``Electric double layer'' (EDL)~\cite{parsons1990a}. From a classical perspective, in perfect metals the electronic charge is localized at the surface and the electric potential is uniform inside the solid, so that the interfacial properties are essentially governed by the ionic densities and electrostatic potential profiles near the interface, which result from the balance between energetic considerations (attraction of counterions to the surface charge and repulsion of co-ions, favoring the build-up of charge) and entropic ones (diffusion leading to uniform concentrations). Of particular interest are of course the charge accumulated on the electrode (and opposite total charge of the interfacial liquid), and the capacitance of the interface, \textit{i.e.} the derivative of the charge with respect to the potential drop across the interface, which can be measured in electrochemical experiments. Voltage also provides a handle on wetting properties, by changing surface free energies and the resulting contact angle: Electrowetting can be used to manipulate liquids, \emph{e.g.} to make actuable lenses. 

While electrostatic interactions play an essential role on the structure, thermodynamics, dynamics and reactivity of electrode-electrolyte interfaces, these properties also crucially depend on the nature of the ions and solvent, as well as that of the metal itself. The finite size of the ions and of solvent molecules results in a layered structure near solid walls, and their ability to form hydrogen bonds (\textit{e.g.} for water) constrains their orientation. Such features, which can be investigated using spectroscopic techniques, depend on the atomic lattice of the metal and a given fluid behaves differently on different faces of the same crystal of a given metal. Finally, many real materials cannot be considered as perfect metals, and the charge and potential distribution within the electrode and their coupling with the liquid must also be taken into account. This effect of the metallic character of the electrode on the properties of the interfacial liquid is a good illustration of the challenge that such interfaces pose for modeling, because by bridging electrons in a solid and ions in a solvent, they are a place where Quantum Chemistry meets Statistical Physics.

On the theoretical side, much progress has been made since the pioneering works of Gouy, Chapman and Stern~\cite{gouy1910a,chapman1913a,stern1924a}. At the same continuous level of description, extensions of the mean-field Poisson-Boltzmann theory have been proposed to capture the effects of electrostatic correlations and excluded volume or solvent polarization~\cite{bazant2011a,goodwin2017a,mceldrew2018a} on the structure and capacitance of the EDL, with a low computational cost compatible with routine use in engineering applications. Even the charging dynamics can be investigated at this level~\cite{bazant2004a,janssen2018a}, even though the effects of ionic correlations or of the coupling with the solvent dynamics are more accurately described by mesoscopic simulations with explicit or implicit ions~\cite{netz2003a,grun2004a,pagonabarraga2010a,lobaskin2016a,asta2019a}. At the other extreme, quantum calculations, usually based on electronic Density Functional Theory (even though Quantum Monte Carlo can now provide even more accurate results \textit{e.g.} on water-carbon interactions~\cite{striolo2016a,brandenburg2019a}), allow to capture the density of states of the metal as well as a detailed description of a few interfacial molecules~\cite{taylor2006a,lautar2020a}. However, their computational cost prevents a fully molecular description of the EDL, and resort to a simplified description of the solvent (polarizable continuum) is the rule rather than the exception. As a result, classical molecular simulations have emerged as a powerful compromise between an atomic description and a computational cost allowing a sufficient sampling of relevant electrolyte configurations.

In the present review, we explore the recent advances on the description and understanding of electrode-electrolyte interfaces with classical molecular simulations. While many applications involve porous electrodes with disordered structures and complex electrolytes such as room temperature ionic liquids, we restrict ourselves to the simpler yet practically relevant and physically rich case of planar interfaces and solvent-based liquids, from pure solvent to water-in-salt-electrolytes (we refer the readers to \textit{e.g.} Refs.~\cite{merlet2012a,merlet2013c,simoncelli2018a,merlet2014a,fedorov2014a,burt2014a,burt2016a,li2017j} on the porous and/or ionic liquid cases).  
In Section~\ref{sec:description}, we discuss the description of electrode-solution interfaces, emphasizing the choice of the models to represent the metallic character of the electrode and its interactions with the electrolyte solution. Section~\ref{sec:simulation} then presents the various strategies to simulate electrochemical systems, with a potential difference between two electrodes, and fundamental issues related to such simulations. Finally, Section~\ref{sec:results} illustrates a selection of properties which can be investigated with such classical molecular simulations, with examples on the capacitance, the interfacial structure and dynamics, electrowetting, as well as steps towards electrochemistry.

\section{HOW TO DESCRIBE ELECTRODE-SOLUTION INTERFACES IN MOLECULAR SIMULATIONS?}
\label{sec:description}

We begin with an overview of models used in classical molecular simulations of electrode-electrolyte interfaces, emphasizing first the description of the electrode (Section~\ref{sec:description:electrode}) and how to capture the electronic response of the metal at this level of description. We then turn in Section~\ref{sec:description:electrolyte} to the description of the electrolyte and the non-electrostatic interactions with the electrode.

\subsection{To be or not to be a metallic electrode}
\label{sec:description:electrode}

\subsubsection{Insulating vs conducting}

The question of how to model metallic electrodes in classical molecular simulations is related to the more general question of the electrostatic response of a medium to an electric charge (ion, or partial charges from molecules), encountered not only in electrochemistry, but in all systems involving interfaces, including \emph{e.g.} biological macromolecules. From a quantum mechanical perspective, electric conduction in a medium is related to the position of the Fermi level relative to the system's energy levels: In a metal, the conduction band is thermally accessible to the electrons, which can be delocalized over the whole material, whereas in an insulator a band gap hinders the conduction of electrons, which remain localized on atomic sites. 

From a classical continuum perspective, an essential feature of these interfaces is a contrast in the polarization response of the various media, quantified by their dielectric constant $\epsilon_r$, ranging from 1 for vacuum to $\approx80$ for liquid water and $\infty$ for a perfect metal. Such a dielectric contrast has strong consequences on the behavior of a charge distribution close to the interface. This is usually expressed in terms of \emph{image charges}. For a sharp flat interface between two media 1 (polar solvent) and 2 (solid wall) with dielectric constants $\epsilon_1$ and $\epsilon_2$, the electrostatic potential arising from a set of charges (ions) ${\bf q}^{ext}=\{q^{ext}_1, \dots, q^{ext}_N\}$ embedded in medium 1 is identical, within this medium, to that arising from a fictitious system in which medium 2 is assigned a dielectric constant $\epsilon_1$ and a set of image charges are placed symmetrically with respect to the boundary (see Figure~\ref{fig:polarization}a below), with magnitudes:
\begin{equation}\label{eq:image-charges}
{\bf q}^{im} = \frac{\epsilon_1 - \epsilon_2}{\epsilon_1 + \epsilon_2} {\bf q}^{ext}
\end{equation}
For an insulating interface such as water-vacuum ($\epsilon_1\gg\epsilon_2$) the image charges are similar to the source charges ${\bf q}_{im} \approx {\bf q}^{ext}$, whereas for a perfect metallic interface ($\epsilon_2\to\infty$) the images have opposite charges ${\bf q}_{im} = - {\bf q}^{ext}$. It follows a radically different electrostatic interaction of a charge with its image: attractive for the metallic case and repulsive for insulators.

\subsubsection{Electrodes in classical molecular simulations}

To model these systems at the atomic scale, one should in principle perform quantum calculations taking into account the electronic density on both the electrode material and the electrolyte. However, this becomes computationally prohibitive for the simulation of large systems over long time scales and alternative approaches have been developed to capture the effect of a metallic electrode on an electrolyte within classical simulations. It should be stressed that the purpose of these models is not to provide an accurate description of the metal and its properties, but rather to reproduce the appropriate boundary conditions for the electrolyte. Within such a simplified description, it becomes possible to sample the configurations explored by the electrolyte at finite temperature and to investigate the physicochemical properties of the system. In this context, two features are particularly important to describe electrochemical interfaces: accounting for the polarization of the metal by the electrolyte, discussed in the following section, and the possibility to accumulate a net charge on the interface, \emph{e.g.} in the presence of an applied voltage between two electrodes. 

\begin{figure}[ht!]
\includegraphics[width=5in]{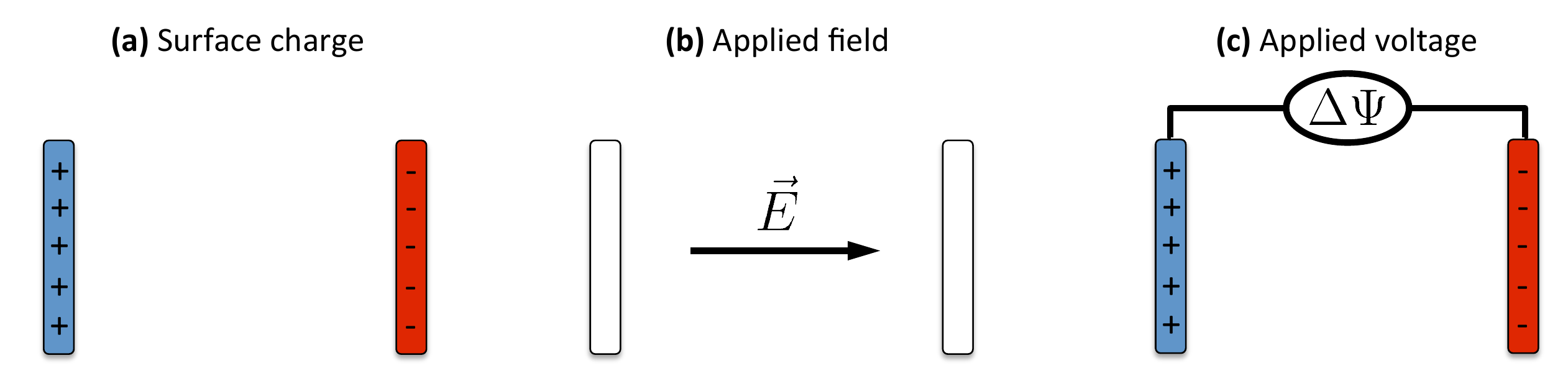}
\caption{A capacitor involves two electrode-electrolyte interfaces and can be modeled with opposite surface charges (a), by applying an external field on the electrolyte (b) or by maintaining the two electrodes at different potentials (c).}
\label{fig:charge}
\end{figure}

This last point is illustrated in Figure~\ref{fig:charge}. A first method to describe charged electrodes is to explicitly put a \emph{constant net charge} on the electrode, using a surface charge on a wall~\cite{torrie_electrical_1980,glosli_molecular_1992,kiyohara_monte_2007,kiyohara_monte_2007-1} or discrete point charges~\cite{van_megen_grand_1980,crozier_molecular-dynamics_2001}. The electroneutrality of the system can be balanced by excess ions in the electrolyte or by an opposite charge on a second electrode. This setup would correspond to a charged pore or an isolated (open circuit) charged capacitor, as shown in panel~\ref{fig:charge}a. From the electrostatic point of view, two oppositely and homogeneously charged wall induce a uniform electric field between them, so that it is (at least in principle) equivalent to directly apply an \emph{external electric field} (panel~\ref{fig:charge}b) on the liquid confined in the electrochemical cell~\cite{lee_molecular_1986,nagy_molecular_1990,hautman_molecular_1989, watanabe_dielectric_1991,rose_adsorption_1993,smith_simulation_1994, zhu_structure_1991,daub_electrowetting_2007}. In a real system, however, the system is rather connected to a voltage generator, which maintains a \emph{constant electric potential difference} between the electrodes and allows the exchange of charge between them~\cite{ilja_siepmann_ordering_1992,siepmann1995a,guymon_simulating_2005,kiyohara_monte_2007,kiyohara_monte_2007-1,reed2007a,pounds2009b,vatamanu_molecular_2009,petersen2012a}, see panel~\ref{fig:charge}c. We will discuss the corresponding simulation setups in Section~\ref{sec:simulation}, and now turn to the representation of polarization of the metal in classical molecular simulations.

\subsubsection{Representing the electronic response of the metal}

\begin{figure}[ht!]
\includegraphics[width=5in]{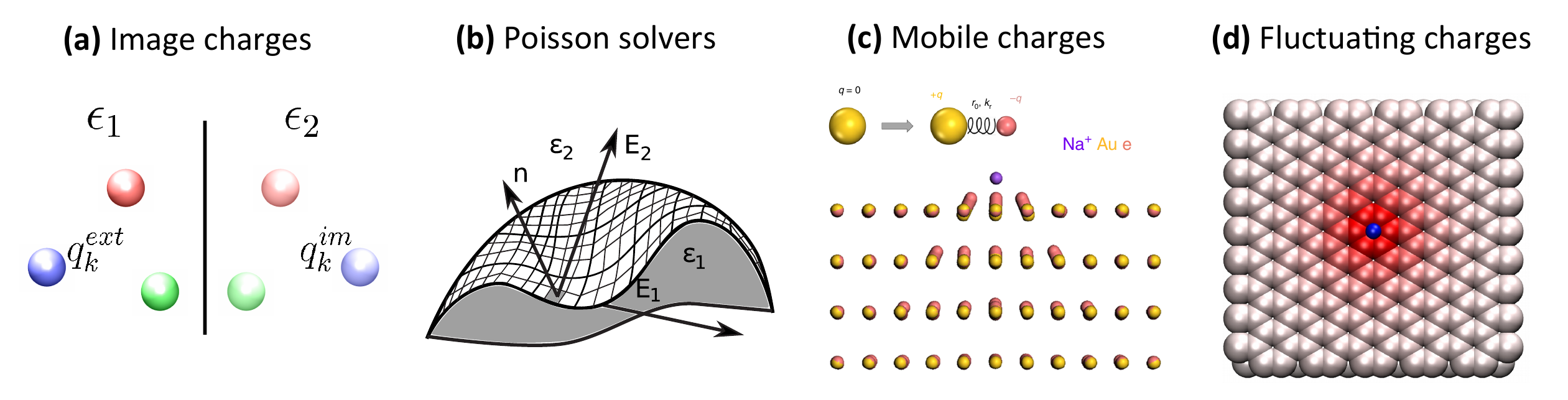}
\caption{Strategies to account for the polarization of the metal by the electrolyte in molecular simuations. (a) Image charges $q^{im}$ inside the solid with permittivity $\epsilon_2$ induced by the electrolyte charges $q^{ext}$ in a medium of permittivity $\epsilon_1$ (see Eq.~\ref{eq:image-charges}). (b) Solving the Poisson equation on a grid to compute the induced (surface) charges. (c) Describing the electronic response with mobile charges: here gold atoms are modeled using opposite charges tethered by a spring, with the cores representing the nuclei (yellow) fixed and the shells (pink) moving in response to the presence of a sodium ion (purple). (d) Describing the electronic response with fixed sites with fluctuating charges: here the charge of atoms in a graphite electrode is distributed inhomogeneously (darker red means more negative) in response to the presence of a cation such as Li$^+$ (blue).
Panel b reproduced from Ref.~\cite{tyagi_iterative_2010}, \emph{J. Chem. Phys.} 2010, {\bf 132}, 154112, with the permission of AIP Publishing; panel c reproduced from Ref.~\cite{geada_insight_2018}, \emph{Nature Commun.} 2018, {\bf 9}, 716, with permission of Springer Nature.
}
\label{fig:polarization}
\end{figure}

Beyond their net charge, a characteristic feature of metallic electrodes is their polarization by the electrolyte (this is even the main effect for neutral electrodes). In the context of molecular simulations, one is primarily concerned with the effect of the metal on the electrolyte and a variety of methods have been proposed. The typical slab geometry of capacitors allows a number of simplifications, including analytical expressions of the electrostatic forces acting on the electrolyte charges and of the electrostatic energy. Even though a uniform charge distribution with varying magnitude can be used to account for an applied external field or voltage, such a description lacks the lateral charge heterogeneities induced by the discrete nature of ions and molecules of the electrolyte. From the electrostatic point of view, an efficient strategy is to impose the proper electrostatic boundary conditions at the surface of the metal using the above-mentioned concept of \emph{image charges}, which can be either explicit or accounted for implicitly in modified Green functions~\cite{torrie_electrical_1982,parsonage_computer_1986,gardner_waterlike_1987,smith_simulation_1994,klapp_monte-carlo_2006,petersen2012a, takae_fluctuations_2015,girotto_simulations_2017}. The use of periodic boundary conditions (see section~\ref{sec:simulation}) also requires special care to compute electrostatic interactions, and efficient algorithms have been developed to deal with the image charges in simulations~\cite{tyagi_icmmm2d_2007,arnold2013a}. Alternatively, the electrostatic problem can also be solved numerically without resorting to image charges: The Induced Charge Computation (ICC) method treats the charge density of the solid as a dynamical variable discretized on a grid and solves the Poisson equation to obtain the induced (surface) charge, as illustrated in Figure~\ref{fig:polarization}b. The dielectric medium is then characterized by a space-dependent dielectric constant $\epsilon({\bf r})$ which can in principle describe arbitrarily shaped interfaces and non-homogeneous media. Based on the variational procedure of Allen \emph{et al.} for solid-electrolyte interfaces~\cite{allen_electrostatic_2001}, extensions using a matrix formulation \cite{boda_computing_2004} or an iterative algorithm (ICC$^*$)~\cite{tyagi_iterative_2010, breitsprecher2015a} were proposed. Another example of this numerical approach to induced charges is the Generalized Minimal Residue (GMRES) method, which provides good performance when used in conjuction with fast Ewald solver~\cite{barros_efficient_2014}.

In order to include the molecular details of the interface (both in the shape and its atomistic nature), one can turn to descriptions based on the electrode atoms, \emph{i.e.} treating the metal at the same level of description as the electrolyte, albeit with dedicated force fields. A first class of such descriptions consists in allowing  \emph{mobile charges} to rearrange in response to the configuration of the electrolyte: This includes core-shell models such as the Drude oscillator~\cite{geada_insight_2018}, with a charge tethered to the electrode atom via a spring (see Figure~\ref{fig:polarization}c), or the rod model~\cite{iori_including_2008,iori_golp_2009,pensado2011a}, in which the auxiliary charge can rotate at a fixed distance around the atom. Such models have the advantage of being easily implemented in standard molecular simulation codes. To ensure a correct adiabatic separation of the charge dynamics and avoid instabilities and/or energy transfers, the mass of the auxiliary charge and the spring strength or rod length should be carefully chosen. A second class of models, which will be described in more detail in the following section, considers instead fixed but \emph{fluctuating charges}, illustrated in Figure~\ref{fig:polarization}d.
One advantage of all these atomistic descriptions compared to those based on image charges is that they are not restricted to slab geometries and can be used to deal with disordered porous electrodes. They capture the polarization effects by rearranging the charge distribution of the metal due to the electric potential created by the electrolyte at each step.
While this is not completely in the scope of the present review, we finally briefly mention attempts at including simplified quantum mechanical treatments of the interface in classical simulations, \emph{e.g.} based on the Jellium model, to represent the spilling of the electronic charge out of the electrode~\cite{schmickler_interphase_1984,shelley_modeling_1997}, or the ``direct dynamics''~\cite{price_molecular_1995,walbran1998a}. Recent developments of mixed quantum/classical (QM/MM) simulations or tight binding approaches might also provide interesting alternatives in the near future.

\subsubsection{Focus on fluctuating charge methods}
\label{sec:fluctuatingcharges}

We now present in some more detail the family of models describing the polarization by assigning charge distributions (typically point charges or Gaussians) to the electrode atoms, with magnitude ${\bf q}=\{q_1,\dots,q_M\}$ treated as additional degrees of freedom which fluctuate in response to the dynamics of ions and molecules of the electrolyte. Early models to account for the polarization of the metal by external charges in fact included both fluctuating charges and induced dipoles at the atomic sites as additional degrees of freedom~\cite{finnis_interaction_1991,finnis_interaction_1995}, but this idea doesn't seem to have been explored much further. The energy of the system is then expressed as a function of the positions ${\bf r}^N=\{{\bf r}_1,\dots,{\bf r}_N\}$ and momenta of all electrolyte atoms (see also section~\ref{sec:description:electrolyte}), and the charges ${\bf q}=\{q_1,\dots,q_M\}$ are determined at each time step either by following an equation of motion, or by responding instantaneously to the motion of the electrolyte in order to impose the electric or electrochemical potential of the electrode atoms (see also Section~\ref{sec:simulation}).

This general idea has resulted in a variety of models, which to some extent share the same quadratic form of the energy as a function of the electrode charges ${\bf q}$, even though the physical meaning of the parameters may differ, and of numerical algorithms to determine the charges at each time step. While fluctuating charge models had already been used for molecules, with the charge equilibration (QE) method or the electronegativity equalization method (EEM)~\cite{nalewajski1984a,mortier_electronegativity-equalization_1986,rappe1991a}, their use to represent a metallic surface was first described by Siepmann and Sprik~\cite{siepmann1995a}, using Gaussian charge distributions on the electrode atoms. This allowed them to simulate a water film near the tip of a scanning tunneling microscope. This model seems to have been proposed again later in Ref.~\cite{guymon_simulating_2005} to model water and ions near copper surfaces. It was then adapted within a Born-Oppenheimer framework to simulate an electrochemical cell by Reed \emph{et al.}~\cite{reed2007a}. The electrostatic energy of the system is written as:
\begin{equation}
\mathcal{U}_{el}({\bf r}^N,{\bf q}) = \frac{ {\bf q}^T {\bf A} {\bf q}}{2} - {\bf q}^T {\bf B}({\bf r}^N) + C({\bf r}^N)
\end{equation}
where the symmetric $M\times M$ matrix ${\bf A}$ depends on the positions of the electrode atoms and the parameters describing the charge distribution on each atom, while the components of the vector ${\bf B}$ are the electrostatic potentials due to the electrolyte on each electrode atom (see Refs~\cite{reed2007a,gingrich2010a} for explicit expressions of ${\bf A}$ and  ${\bf B}$ in the particular case of Gaussian charge distributions with 2D Ewald summation), and the scalar $C$ corresponds to electrostatic interactions within the electrolyte. The set of electrostatic potentials on each electrode atom is given by the gradient of $\mathcal{U}_{el}$ with respect to ${\bf q}^T $,
\begin{equation}
\label{eq:elecpot}
\frac{\partial \mathcal{U}_{el}({\bf r}^N,{\bf q})}{\partial {\bf q}^T } = {\bf A} {\bf q} - {\bf B}
\; ,
\end{equation}
and depends on the positions of the electrode atoms and on the electrolyte configuration. As a result, the set of charges satisfying the constraint of fixed electrostatic potentials ${\bf \Psi}=\{\Psi_1,\dots,\Psi_M\}$ for each atom, typically the same value for all atoms belonging to a given electrode and a difference $\Delta\Psi$ between the values for both electrodes, is given by ${\bf q}={\bf A}^{-1} \left( {\bf B} + {\bf \Psi}\right)$. In practice, if the matrix ${\bf A}$ can be inverted numerically (this has to be done only once if the electrode atoms do not move), the charges are computed at each step by a simple matrix-vector multiplication. Other methods are possible, such as finding the charges by minimizing $\mathcal{U}_{el}- {\bf q}^T {\bf \Psi}$ numerically, \emph{e.g.} with the conjugate gradient method, or treating the fixed potential as a holonomic constraint~\cite{coretti_mass-zero_2020}. 

The expression of the energy is similar in the case of the QE/EEM method, which was used to investigate electrochemical interfaces~\cite{streitz_electrostatic_1994,onofrio2015b,onofrio2015a,liang_applied_2018,nakano_chemical_2019,buraschi2020a}, but the expressions of the matrix ${\bf A}$ and vector ${\bf B}$ involve terms such as electronegativities and chemical hardnesses, instead of the purely electrostatic picture described above. It is also related to the split charge equilibration approach, which includes bond-specific terms in the energy~\cite{nistor_dielectric_2009}. Finally, Pastewka \emph{et al.} pushed the concept of using fluctuating charges as a proxy for a quantum description even further, by also parametrizing the band-structure energy, which appears in the tight-binding approximation with self-consistent charges to describe the terms beyond the electrostatic interactions between them, as a function of ${\bf q}$~\cite{pastewka_charge-transfer_2011}. This allowed a purely classical description of various carbon electrodes, taking into account to some extent their different band structures. The potential of such a promising strategy doesn't seem to have been much exploited so far.

\subsection{It takes two to tango}
\label{sec:description:electrolyte}

\subsubsection{Description of the electrolyte}

On the other side of the interface, the electrolyte can be represented using various levels of sophistication. In molecular dynamics simulations, two main families of models are generally employed~\cite{salanne2015a}. The first one consists in all-atom models, which include all the atomic details of the molecules. Intramolecular interactions involve a set of two, three and four-body terms that are optimized to reproduce all the bonds, angles and dihedrals of the molecule accurately. In some cases, constraints are used to cancel some degrees of freedom, generally the bonding terms involving hydrogen atoms in order to allow the use of larger timesteps in the simulations (the typical timestep of all-atom force fields-based MD simulations is 1~fs). 

The second family of models represents the molecules using ``coarse grains''. The so-called grains are interaction sites representing several atoms. The level of coarsening may be tuned, resulting in a compromise between computational cost (due to the reduced number of interacting sites and the use of larger timesteps -- up to 5~fs) and accuracy. In the case of electrochemical interfaces, coarse-grained models generally include between 4 and 10 atoms per grains~\cite{merlet2013a}, resulting in a decrease of the number of interaction sites by one order of magnitude compared to all-atom models (note that an alternative consists in using ``united-atom'' models in which only the hydrogen atoms are merged with heteroatoms to which they are bonded). As in the latter case, coarse-grained models may introduce some intramolecular potentials, although their physical roots are then more difficult to establish, or use constraints to fix the geometry of the molecule, treated as a rigid body in the simulations. 

In both families of models, the intermolecular interactions will then determine most of the liquid properties. Many analytical expressions are used in order to account for the short-range repulsion between the electronic clouds, the dispersion effects and the electrostatic interactions. Concerning electrostatics, all the models employ partial charges which are distributed among the interaction sites of the molecules, which interact between them and with the electrode charges. In a few cases, polarization effects are included by adding either charge-on-a-spring (Drude oscillator model) or induced dipoles on the atomic sites~\cite{tazi2010a,bedrov2019a,park_interference_2020}. These simulations are more costly from the computational point of view since they involve additional degrees of freedom that have to be either propagated or determined by solving a set of self-consistent equations (they have a many-body character), but they should in principle be more accurate since the electrostatics interactions are of primary importance at the interface.  An alternative is to use rescaled charges in order to mimic these polarization effects~\cite{lebreton2020a}.

\subsubsection{Electrode-electrolyte: non-electrostatic interactions}

The electrostatic interactions between the electrode and the electrolyte ions and molecules follow from their respective descriptions as charge and dipole distributions, discussed above. Another crucial aspect is to account for the repulsive and dispersive intermolecular interactions between the electrolyte atoms and the electrode surface. In the earliest studies, the surface was materialized by a structure-less wall, that took the form of a hard wall, a purely repulsive potential or different flavors of one-dimensional Lennard-Jones potentials such as the Steele potential~\cite{lee_molecular_1986, russier_adsorption_1987, parsonage_computer_1986, hautman_molecular_1989, kiyohara_monte_2007}. These descriptions are computationally less demanding but lack important molecular features close to the interface, that give rise \emph{e.g.} to specific adsorption sites or templating effects. Atomic descriptions of the electrodes are therefore employed, using a finite number of discrete atoms, in order to account for different geometries and crystal structures~\cite{siepmann1995a, geada_insight_2018, daub_electrowetting_2007, iori_golp_2009}. Usual intermolecular potentials are employed such as Lennard-Jones and Born-Mayer potentials. In most of these studies, the non-electrostatic electrode-electrode interactions are neglected, because the individual electrode atoms are immobile, with a few exceptions where the electrode structure is rigid but can translate as to mimic a piston at constant pressure, as done in Ref.~\cite{coretti_mass-zero_2020}; in only a few cases, the electrodes atoms are free to move and interact with a harmonic potential~\cite{spohr_molecular_1986}. It should be highlighted that the inclusion of explicit atoms for the intermolecular part of the interactions does not necessarily imply an atomistic treatment of the electrostatic part and vice versa. Mixing atomistic and structure-less descriptions however poses the problem of where to locate the interface (in a continuum picture) with respect to the atomic positions. Intermediate approaches have also been used, which formulate the interaction potential as a corrugated potential, avoiding the calculation of pair terms but reproducing the local roughness of the substrate \cite{spohr_computer_1989}. 

As usual, the parametrization of these non-electrostatic interactions, which may lead or not to good predictions depending on how the electrostatic part is described, is crucial. While some force fields were fitted on {\it ab-initio} calculations or to reproduce experimental data, in the absence of reliable data the choices are often based on the availability of parameters from other studies. For carbon electrodes, due to the lack of accurate experimental atomic-scale data, no special optimization was made and generic parameters for carbon were then employed~\cite{merlet2013a}. However, it is worth noting that accurate quantum Monte-Carlo reference data were reported for interfaces between water and graphene or carbon nanotubes~\cite{alhamdani2017a}, that could be used to develop new classical force fields in the near future. Comparatively, more efforts were put in the representation of metal surfaces. For example, Heinz {\it et al.} have proposed a systematic parameterization of Lennard-Jones potentials for several face-centered cubic metals, that were shown to reproduce a few experimental data for interfaces with water, such as surface tensions~\cite{heinz2008a}. When used in combination with a core-shell model to represent the polarization of the metal, it was necessary to reparametrize the Lennard-Jones potentials as well~\cite{geada_insight_2018}. In the case of platinum, Siepmann and Sprik showed that using a three-body function was necessary to push water molecules on top of metal atoms to represent chemisorption effects~\cite{siepmann1995a}. An alternative approach was recently proposed to include these effects through an attractive two-body Gaussian potential~\cite{clabaut2020a}. We will also come back to reactive force fields to capture the breaking and formation of bonds in Section~\ref{sec:electrochemistry}.

\section{HOW TO SIMULATE POLARIZED ELECTRODE-SOLUTION INTERFACES}
\label{sec:simulation}

Beyond the choice of the force field to describe the metallic electrode, the electrolyte and the interactions between them, a molecular simulation requires the definition of the simulated system and an algorithm to sample its configurations in order to compute physical properties. In this section, we explore more specifically the available options to simulate an electrochemical cell with an applied potential difference between two electrodes, as well as issues related to the sampling of the corresponding microscopic configurations. Even though the discussion below applies more generally, we illustrate the various points with a description of the electrode based on fluctuating charges. In addition, we will not discuss in detail the possibilities offered by existing open source softwares for the classical molecular simulation of electrochemical interfaces with the methods described in the present review, but refer the reader to the corresponding descriptions of \emph{e.g.} \emph{Metalwalls}~\cite{Marin-Lafleche2020}, which is dedicated to the simulation of such interfaces, or more generic simulation packages allowing such simulations (possibly with an open source modification not provided in the standard distribution) such as \emph{LAMMPS}~\cite{plimpton1995a}, \emph{OpenMM}~\cite{eastman_openmm_2017} or \emph{ESPRESSO}~\cite{weik_espresso_2019} (see~\cite{AllCodes} for links).

\subsection{Increase tension (between electrodes) or handle (the electrolyte) with care}

\subsubsection{Simulation setup}

In a typical electrochemical system, two electrodes are separated by a slab of liquid electrolyte with all dimensions along the surfaces and between them much larger than molecular ones. In order to ensure that the two interfaces do not interfere (even though the total charge accumulated on both electrodes are of course correlated, see also Section~\ref{sec:electroneutrality}), the distance between the electrodes must be large compared to the screening length. In experiments, one also introduces a separator to prevent contact (hence short-circuit) between the electrodes. The typical dimensions of systems that can be simulated with classical molecular simulations is in the range of a few to a few tens of nanometers. This requires a simplified description of the real device and limits the range of physical systems that can be simulated in a meaningful way. In particular, one always neglects the presence of the separator, which is a safe assumption provided that the distance between the electrodes remains large compared to the screening length. This in turn sets a lower bound on the ionic concentrations that can be considered -- typically $\sim$0.1~mol~L$^{-1}$, a limit which also emerges from the constraint of having enough ions in the simulation box to ensure a good sampling of the phase space. 

\begin{figure}[ht!]
\includegraphics[width=4in]{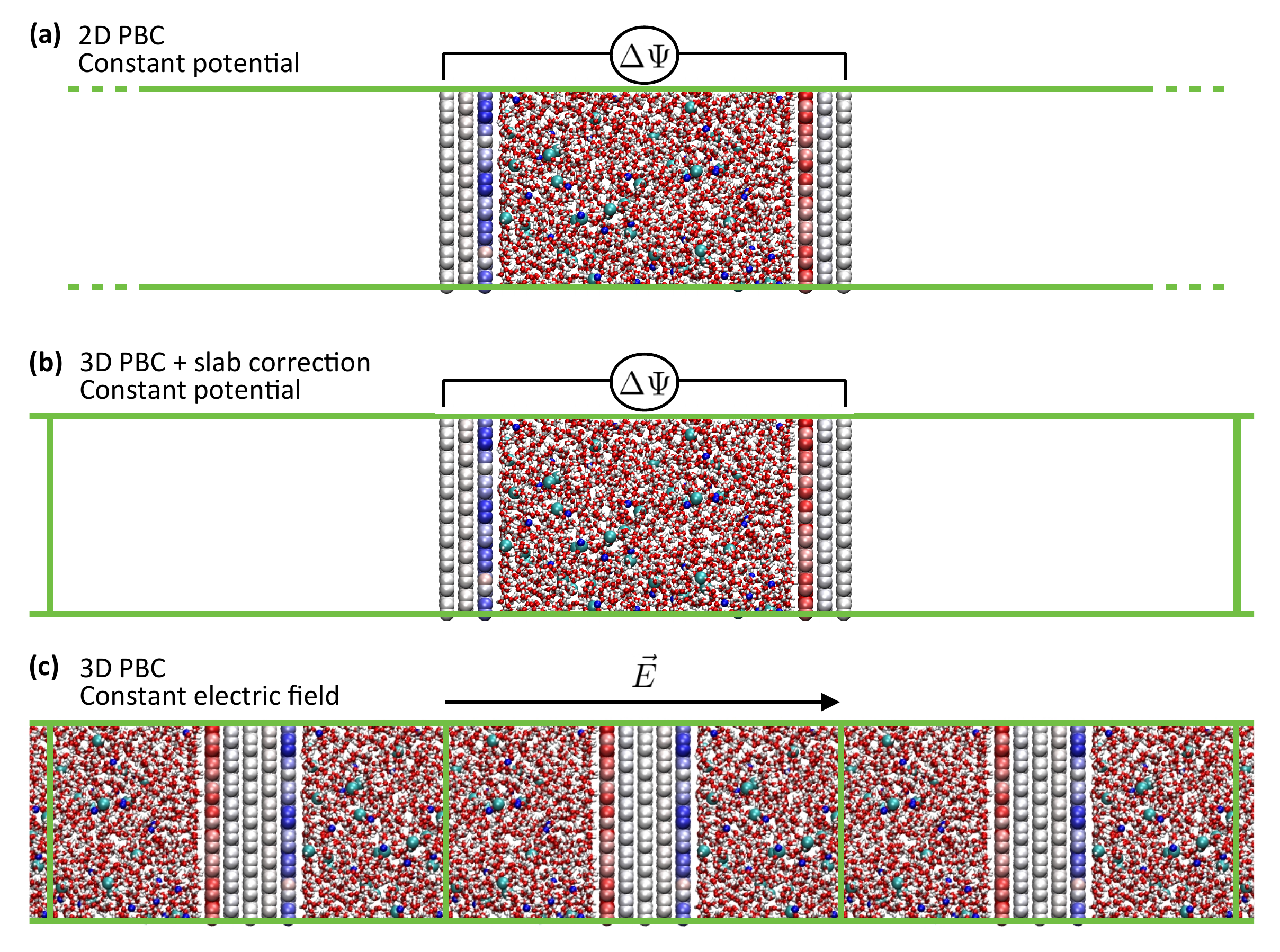}
\caption{
Simulation setups to simulate a capacitor consisting of graphite electrodes separated by a 1~mol~L$^{-1}$ aqueous NaCl solution under a voltage $\Delta\Psi=2$~V; the color of the electrode atoms reflects their instantaneous charge. (a) Periodic boundary conditions (PBC) are applied only in the two directions along the surfaces, but not in the direction perpendicular to the electrodes, and the two electrodes are maintained at different potentials, corresponding to a voltage $\Delta\Psi$. (b) PBC are applied in all directions, but the simulation box also includes vacuum (here the length of the box in this direction is three times larger than the electrochemical cell) and a ``slab'' correction is applied when computing electrostatic interactions; as in the previous case, the two electrodes are maintained at different potentials. (c) PBC are applied in all directions, without vacuum; all electrode atoms are maintained at the same potential and the capacitor is charged by applying an electric field to the electrolyte such that $\Delta\Psi=-EL_z$, with $L_z$ the box dimension in the direction perpendicular to the electrode.
}
\label{fig:setups}
\end{figure}

Last but not least, the small number of atoms in the system (10$^3$-10$^5$) compared to a real device requires, as always in molecular simulation of condensed matter, the use of periodic boundary conditions (PBC). Based on the above discussion, it is natural to consider PBC in the two directions along the surfaces, but not in the direction perpendicular to the electrodes, as illustrated in Figure~\ref{fig:setups}a. This requires in particular computing long-range electrostatic interactions with dedicated approaches, such as 2D Ewald summation, also taking into account the description of electrode atoms by Gaussian charge distributions instead of point charges if needed~\cite{reed2007a,gingrich2010a}. An alternative approach, more easily implemented in standard packages, is to mimick 2D PBC by considering PBC in all directions, but adding vacuum to the simulation box, as illustrated in Figure~\ref{fig:setups}b. In that case, it is necessary to apply a ``slab'' correction when computing electrostatic interactions~\cite{yeh1999a}.

\subsubsection{Constant-voltage or finite field}

The simulation of an electrochemical cell also requires a way to impose voltage between the two electrodes. The most straightforward way to achieve this, with descriptions allowing to set the potential of electrode atoms such as methods based on fluctuating charges, is to impose different values of the potential for atoms belonging to the two electrodes, \emph{i.e.} one value for each electrode with a difference $\Delta\Psi$ between the two electrodes, as illustrated in panels~\ref{fig:setups}a and~\ref{fig:setups}b for the 2D and 3D-slab PBC cases, respectively. Recently, an alternative approach has been proposed to allow the use of 3D PBC without the need to introduce vacuum, hence decrease the computational cost, and facilitate the implementation in standard simulation packages. The method introduced by Dufils \emph{et al.}~\cite{dufils_simulating_2019}, illustrated in Figure~\ref{fig:setups}c, consists in applying a finite electric field to the electrolyte via an extended Hamiltonian. Such finite field methods, developed in the framework of the modern theory of polarization~\cite{stengel2009b}, had first been adapted to investigate EDLs near charged or polar insulator-electrolyte interfaces~\cite{zhang2016f,sayer2017a}. By imposing a constant potential on all the electrode atoms (there is a single electrode in this setup) using the fluctuating charge method and an electric field on the electrolyte such that $\Delta\Psi=-EL_z$, with $L_z$ the box dimension in the direction perpendicular to the electrode, one can charge the capacitor by creating two EDLs, one on each side of the electrode. If the width of the electrode is sufficient, one recovers the same results (in particular for the induced charges on the electrode or the structure of the interfacial electrolyte) as in the 2D periodic case of Figure~\ref{fig:setups}a, with a reduced computational cost. A further benefit of this finite field approach is that it could easily be used with other methods to describe the metal (it had already been used in 3D PBC with the ICC method~\cite{arnold2013a}) with explicit electrode atoms, \emph{e.g.} with the core-shell model of Ref.~\cite{geada_insight_2018}, which capture the polarization of the metal but does not provide a handle to control the potential of the electrodes (hence the voltage between them), or even in \emph{ab initio} simulations~\cite{bonnet_first-principles_2012}.


\subsection{Enough is enough? Sampling configurations}

While the previous section presented the difficulties related to the description of the system, we now turn to the theoretical and numerical challenges pertaining to the sampling of microscopic configurations of electrode-electrolyte interfaces. Molecular simulations are inherently related to Statistical Mechanics and the fundamental problem of sampling configurations from a thermodynamic ensemble, from which physical properties can be computed as averages. Even though most molecular simulations with constant-potential electrodes are performed in the canonical ensemble (constant number of electrolyte ions and molecules, $N$, volume $V$ and temperature $T$), one can in principle also consider a constant normal pressure on the electrodes (at least with rigid electrodes acting as pistons), or even grand-canonical (fixed chemical potential) and Gibbs ensemble (exchange between two systems) simulations. In the latter cases, which involve the exchange of particles, it might be necessary to resort to Monte Carlo simulations, as done \emph{e.g.} in Refs.~\cite{kiyohara_monte_2007,kiyohara_monte_2007-1,punnathanam_gibbs-ensemble_2014,stenberg_grand_2020} instead of molecular dynamics more commonly employed in this field. We focus here on three separate though related aspects, which apply generally to all the above situations, but limiting ourselves to the more common canonical ensemble: the issue of global electroneutrality, the statistical mechanics of the constant-potential ensemble, and importance sampling.


\subsubsection{Electroneutrality}
\label{sec:electroneutrality}

When studying an electrochemical system with electrodes held at a constant potential difference by a generator at a classical molecular dynamics level, the main focus and interest is rather on the electrolyte properties.  Importantly, the details of the device that applies a potential difference are not taken into account. Instead the models consider an open system, very much like in a grand-canonical ensemble, exchanging charge (instead of particles) with a reservoir. In order to properly separate the electrochemical cell from the electric potential generator, the interaction between them should be negligible. This requires both subsystems to be electroneutral so as to cancel the leading term in the long-range electrostatic interaction. Although in the real device there are fluctuations and the electroneutrality of the cell is only verified on average, it should be enforced at a microscopic level in molecular simulations not explicitly including the charge reservoir. This can be achieved easily in descriptions with fluctuating charges by adding a Lagrange multiplier to enforce this constraint when computing the charges of the electrode atom. This multiplier corresponds to a potential shift applied to both electrodes simultaneously, while keeping the potential difference unchanged~\cite{haskins2016a,scalfi2020a,coretti_mass-zero_2020}.

\subsubsection{Statistical mechanics of the constant-potential ensemble}
\label{sec:statmech}

The natural ensemble associated with the above-mentioned simulations is the canonical constant-potential ensemble, characterized by the total number of atoms $N$, the volume $V$, the temperature $T$ and the applied potential $\bf \Psi$ (more precisely, the set of potentials imposed no each electrode atom, usually one value per electrode). The charge distribution on the electrode atoms is then the conjugate variable of the applied potential and its fluctuations contain useful information. The aim of molecular simulations is to sample configurations from this ensemble in order to compute meaningful statistical averages. Along with approximations in the choice of force fields and the level of description of the quantum degrees of freedom, there are approximations related to how the chosen algorithm samples this thermodynamic ensemble.
Using constant charges on electrode atoms ignores any charge fluctuation around the mean. One can use external inputs or post-processing to link the total electrode charge to an equivalent applied potential, but the configurations are not sampled according to their weight in the constant-potential ensemble.

Even though in principle one could also use Monte Carlo algorithms to sample the electrolyte configurations and electrode charges, to date most simulations using fluctuating charge models used molecular dynamics. In addition, except in earlier studies using Car-Parrinello (CP) dynamics with a fictitious mass for the additional degrees of freedom~\cite{siepmann1995a}, the vast majority of simulations employing fluctuating charge models rely instead on Born-Oppenheimer (BO) dynamics, by assuming a separation of time scales between the electronic and nuclear degrees of freedom~\cite{reed2007a}. This allows using larger time steps, but reduces the full phase space to that of electrolyte configurations, since to any such configuration corresponds a single set of electrode atom charges ${\bf q}^*$. A thorough discussion of the statistical mechanics of the constant-potential ensemble can be found in Ref.~\cite{scalfi2020a}, where we clarified in particular which observables can be computed exactly using BO dynamics on the charges. This is \emph{e.g.} the case of the average total charge of an electrode $\left\langle Q_{tot} \right\rangle=\left\langle Q_{tot}^* \right\rangle$. In contrast, the variance of the total charge, $\left\langle\delta Q_{tot}^2\right\rangle$, which is related to the differential capacitance by~\cite{kiyohara_monte_2007,limmer2013a}:
\begin{equation}
\label{eq:fluctuationdissipation}
C_\mathrm{diff} = \frac{\partial \left\langle Q_{tot} \right\rangle }{\partial \Delta\Psi }= \beta\left\langle\delta Q_{tot}^2\right\rangle
\end{equation}
with $\Delta\Psi$ the voltage between the two electrodes and $1/\beta=k_BT$ the thermal energy, includes a contribution from the suppressed thermal charge fluctuations. The differential capacitance is then given by
\begin{eqnarray}
\label{eq:fluctuationdissipation2}
C_\mathrm{diff} &
= C_\mathrm{diff}^\mathrm{electrolyte} + C_\mathrm{diff}^\mathrm{empty}
= \beta\left\langle\delta Q_{tot}^{*2}\right\rangle + C_\mathrm{diff}^\mathrm{empty}
\end{eqnarray}
where $\left\langle\delta Q_{tot}^{*2}\right\rangle$ is the variance sampled within BO dynamics and $C_\mathrm{diff}^\mathrm{empty}$, for which an explicit expression can be found in Ref.~\cite{scalfi2020a}, corresponds to the empty capacitor (\emph{i.e.} in the absence of thermal fluctuations of the electrolyte).

\subsubsection{Importance sampling}
\label{sec:importancesampling}

As usual in molecular simulations, one may also face sampling issues due to physical processes that result in long time scales not accessible by straightforward simulations. This includes \emph{e.g.} slow transport within the electrolyte, or adsorption/desorption at the interface, or even phase transitions. In that case, one can resort to dedicated approaches, such as umbrella sampling to compute free energies, or transition path sampling to compute rates and analyze mechanisms. Examples of studies  in the context of electrolyte interfaces, illustrated in Section~\ref{sec:dynamics}, involve collective variables such as the distance of an ion to the surface or between two ions to investigate their adsorption/desorption and pair formation/dissociation, or the number of water molecules in a probe volume, to investigate the hydrophilic/phobic behavior of the interface.

An importance sampling approach more specific to electrode-electrolyte interfaces was proposed in Ref.~\cite{limmer2013a}. By adapting the standard histogram reweighting approach to the case of constant-potential simulations, it is possible to combine simulations performed at different potentials in order to optimally compute the properties of the system as a function of voltage, including for voltages for which no simulations are performed. The only requirement is an overlap of the distributions of the total charge $P(Q_{tot})$, or the joint distribution $P(Q_{tot},\textit{p})$, with $\textit{p}$ a property of interest such as density profiles or orientational distributions, between the simulations at various voltages. This approach has been used for example in the study illustrated in Section~\ref{sec:structure}. Even though to the best of our knowledge it has never been done to date, one could use the same idea to derive a parallel tempering algorithm, exchanging replicas between simulations at different potentials (instead of temperature in the original method) to enhance the sampling of phase space. Histogram reweighting has also been used recently in Ref.~\cite{takahashi_polarizable_2020} to sample the so-called vertical energy gap related to electron transfer reactions (see Section~\ref{sec:electrochemistry}). Other strategies not requiring an explicit biais on the collective variable of interest, but taking advantage of statistical tools, include indirect umbrella sampling to reweight configurations using a bias on an auxiliary collective variable~\cite{patel_fluctuations_2010}, which was applied in Ref.~\cite{kattirtzi2017a} to the Madelung potential experienced by ions at an electrochemical interface (see Section~\ref{sec:dynamics}).

\section{WHAT CAN WE LEARN FROM MOLECULAR SIMULATIONS OF ELECTRODE-SOLUTION INTERFACES?}
\label{sec:results}

Once a description of the system and a sampling strategy are chosen, one can perform classical molecular simulations to compute observable properties. In this section, we provide a selection (hence by no means exhaustive list) of illustrations of such properties typical of electrode-electrolyte interfaces: capacitance, interfacial structure and dynamics, electrowetting and an opening towards electrochemistry. All examples have in common the use of a fluctuating charge model to describe the electrode.

\subsection{Capacitance}
\label{sec:capacitance}

We begin by considering one of the most relevant property for practical applications, namely the capacitance. More precisely, one generally applies a voltage $\Delta\Psi$ between two metallic electrodes separated by an electrolyte, and the charge $\pm Q_{tot}$ on both electrodes, which fluctuates in response to the thermal motion of the liquid (see Section~\ref{sec:statmech}), allows the definition of an integral capacitance $C_\mathrm{int}=\left\langle Q_{tot} \right\rangle / \Delta\Psi$ and a differential capacitance $C_\mathrm{diff}= \partial \left\langle Q_{tot} \right\rangle / \partial \Delta\Psi$.
Both quantities coincide only when the differential capacitance does not depend on voltage, \emph{i.e.} when the response to applied voltage is linear. Such a linear behavior is expected from continuum electrostatics at least for (a) polar liquids behaving as pure dielectric media with relative permittivity $\epsilon_r$; in that case $C=\epsilon_0\epsilon_r \mathcal{A}/L$, with $L$ the distance between the electrodes and $\mathcal{A}$ their surface area and (b) dilute electrolyte solutions; in that case $C=\epsilon_0\epsilon_r \mathcal{A}/\lambda_D$, with the Debye screening length $\lambda_D= \left( \frac{e^2}{\epsilon_0 \epsilon_r k_BT} \sum_i c_i z_i^2 \right)^{-1/2}$, where $c_i$ and $z_i$ are the ionic concentrations and valencies, respectively and the sum runs over ionic species $i$. In practice, however, the organization of the molecular solvent and the distribution of ions at the interface do not exactly follow the assumptions leading to these simple expressions, and molecular simulations allow to evaluate $C_\mathrm{int}$ and $C_\mathrm{diff}$, as well as to correlate their evolution as a function voltage with structural changes in the interfacial fluid.

\begin{figure}[ht!]
\includegraphics[width=5in]{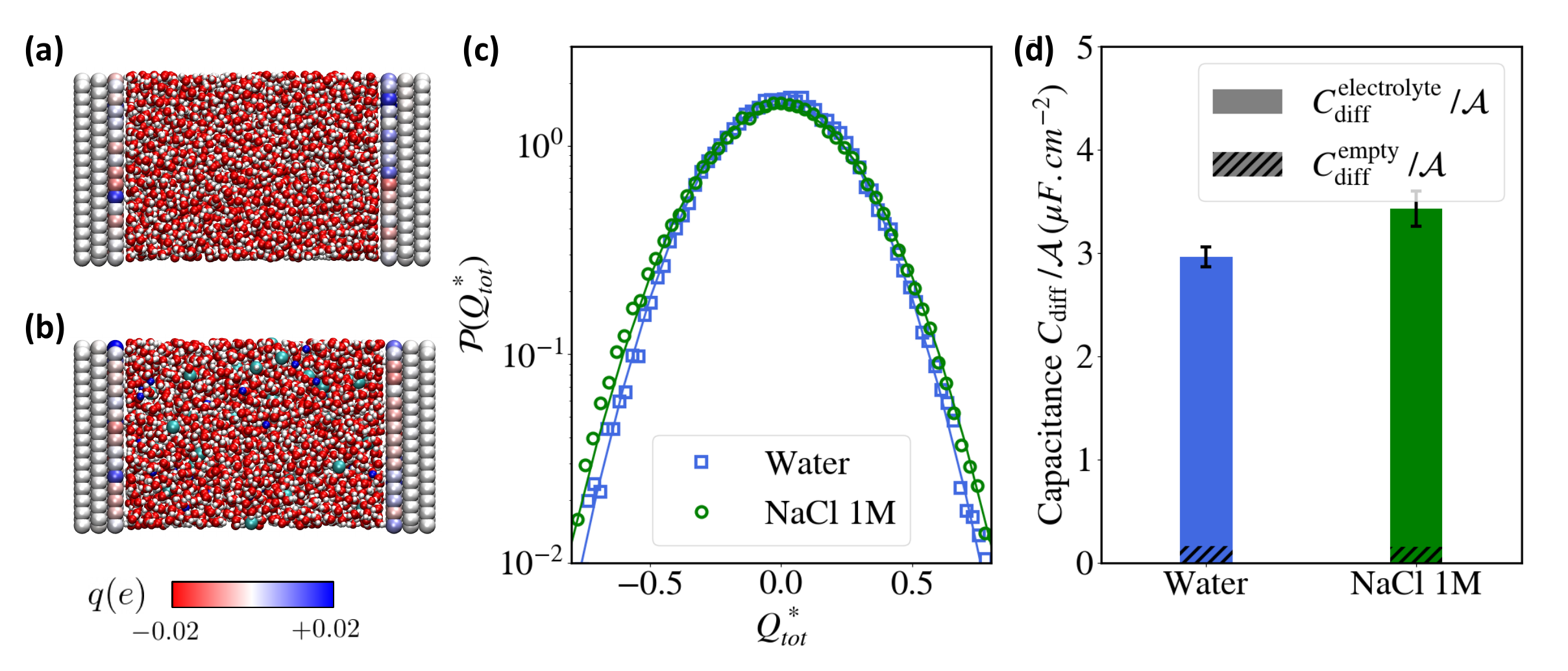}
\caption{
Capacitors consisting of graphite electrodes separated by pure water (a) and a 1M aqueous NaCl solution (b) under a voltage $\Delta\Psi=0$~V; the color bar indicates the instantaneous charge of the electrode atoms. (c) Distribution of the total charge of the electrode $Q_{tot}^*$ for both systems, computed within Born-Oppenheimer dynamics simulations; solid lines are normalized Gaussian distributions with the corresponding standard deviation. (d) Contributions to the differential capacitance per unit area $C_\mathrm{diff}/\mathcal{A}$ at 0~V: from the empty capacitor, $C_\mathrm{diff}^\mathrm{empty}$, and from the charge fluctuations induced by the electrolyte, $C_\mathrm{diff}^\mathrm{electrolyte}=\beta\left\langle\delta Q_{tot}^{*2}\right\rangle$ (see Eq.~\ref{eq:fluctuationdissipation2}); the latter corresponds to the variance of the distributions of panel (c).
Panels a, b and d adapted from Ref.~\cite{scalfi2020a}, \emph{Phys. Chem. Chem. Phys.}, 2020, {\bf 22}, 10480 by permission of the PCCP Owner Societies.
}
\label{fig:capacitance}
\end{figure}

In practice, one could compute the average charge $\left\langle Q_{tot} \right\rangle$ as a function of voltage and perform a numerical derivative to obtain the differential capacitance. A much more efficient and accurate approach is to use the fluctuation-dissipation relation Eq.~\ref{eq:fluctuationdissipation2}, introduced in Ref.~\cite{limmer2013a} and completed in Ref.~\cite{scalfi2020a}, which provides $C_\mathrm{diff}$ from the variance of the charge distribution in a single simulation. Figure~\ref{fig:capacitance} illustrates this method with results from Scalfi \emph{et al.} for two capacitors consisting of graphite electrodes separated by pure water or a 1M aqueous NaCl solution (see panels~\ref{fig:capacitance}a and ~\ref{fig:capacitance}b, respectively), under a voltage $\Delta\Psi=1$~V~\cite{scalfi2020a}. The variance of the distribution $P(Q_{tot}^*)$ of the total charge computed in Born-Oppenheimer dynamics simulation in the constant-potential ensemble (see section~\ref{sec:statmech}), shown in panel~\ref{fig:capacitance}c, provides the contribution of the electrolyte, $C_\mathrm{diff}^\mathrm{electrolyte}=\beta\left\langle\delta Q_{tot}^{*2}\right\rangle$, to the differential capacitance, reported in panel~\ref{fig:capacitance}d together with that of the empty capacitor, $C_\mathrm{diff}^\mathrm{empty}$. The latter is small compared to that arising from the charge induced by the presence of the liquid and its thermal fluctuations, consistently with the observation of Haskins and Lawson for ionic liquids~\cite{haskins2016a} -- and with the fact that it had escaped the attention of the community until recently. For pure water, the above-mentioned continuum prediction for the capacitance, taking the permittivity of the SPC/E water model used in this work, results in a value 3-4 times larger than the one obtained reported in Figure~\ref{fig:capacitance}d, because it fails to account for the efficient screening of the field by the first layers of interfacial molecules~\cite{willard2009a,bonthuis2012a,jeanmairet2019b}. Adding salt increases the capacitance compared to pure water, but this increase is moderate for the considered concentration and inter-electrode distance. As for the pure water case, the above-mentioned continuum prediction is not accurate, as expected for Debye-H\"uckel theory at such a high concentration. The shortcomings of such continuous descriptions to estimate the capacitance (both for pure solvent and electrolyte solutions) reflect their limitations to account for the detailed organization of the interfacial fluid, which can also be investigated by molecular simulation, as discussed in the next section.

\subsection{Interfacial structure}
\label{sec:structure}

One of the most straightforward yet valuable information provided by molecular simulation is the microscopic structure of the system. In the case of planar electrode-electrolyte interfaces, density profiles as a function of the position $z$ in the direction normal to the interface can be computed from the trajectories as an ensemble average as $\rho_\alpha(z)=\left\langle \sum_{i\in\alpha} \delta(z_i -z) \right\rangle$, with $\delta$ the Dirac delta function and where the sum runs over atoms $i$ of type $\alpha$. Similar definitions can be used to compute (1D, 2D or 3D) number, charge or polarization density profiles, distributions of molecular orientation, radial distribution functions, coordination numbers, etc. From the charge density profiles, one can also obtain electrostatic potential profiles by integrating the 1D Poisson equation, to make the link with the interfacial capacitance (in the presence of ions, using the potential drop across each interface) or the dielectric permittivity of the liquid (in the absence of ions, using the slope of the potential in the bulk region). The insights from molecular simulations are particularly valuable, since they capture explicitly the effects of the discrete nature of ions and solvent molecules, which lead in particular to their layering near the planar electrode, and of their complex interactions (steric, electrostatic, dispersion, ...). This allows a detailed investigation of the composition and organization of the interfacial liquids and their evolution as a function of applied voltage. The results can also serve as reference data for simpler theories, such as the ones mentioned in the introduction.

\begin{figure}[ht!]
\includegraphics[width=5in]{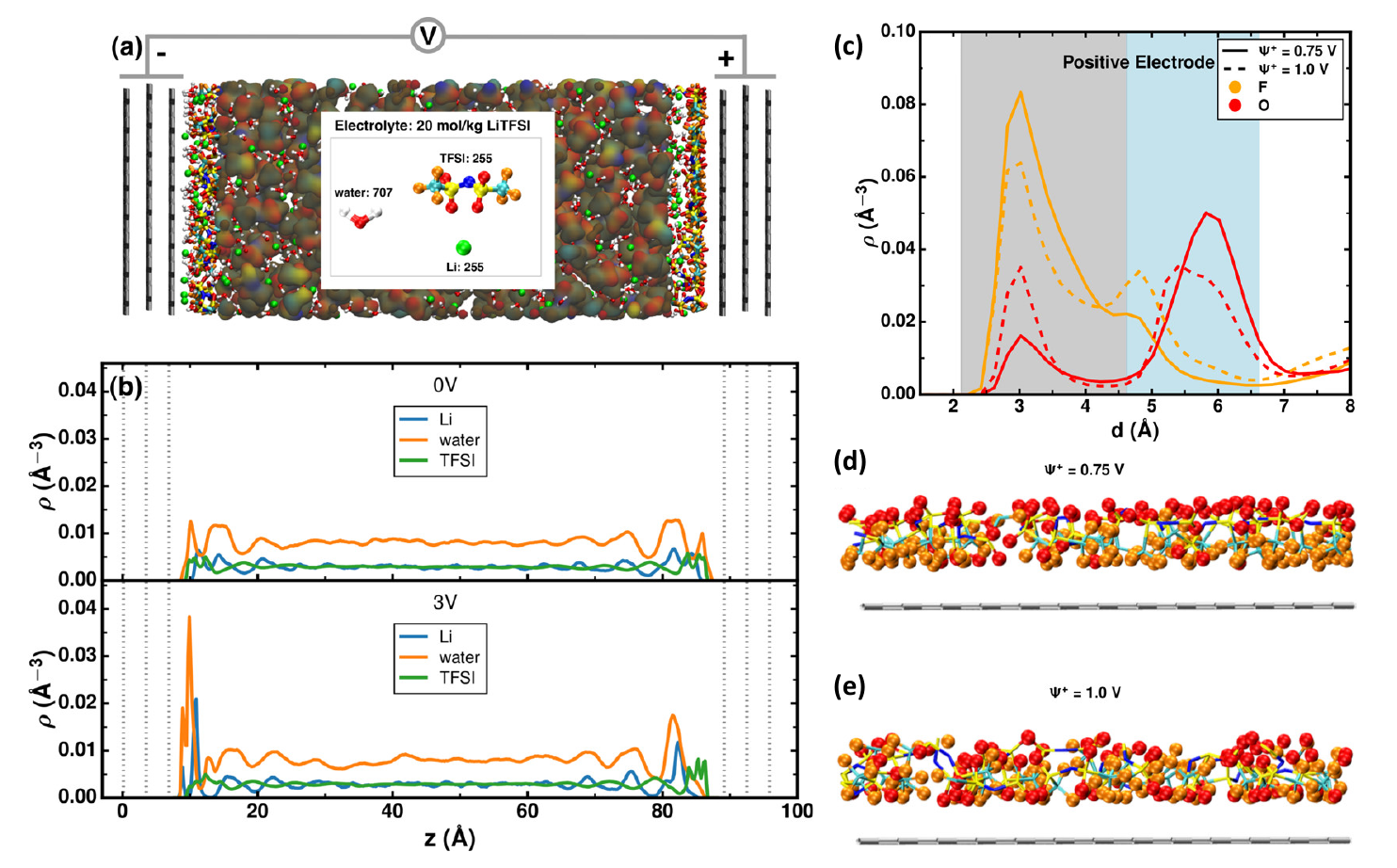}
\caption{
(a) Capacitor consisting of graphite electrodes separated by a water-in-salt-electrolyte (20 mol/kg aqueous LiTFSI solution). 
An all-atom model is used, with H (white), O (red), Li (green), F (orange), C (cyan), S (yellow), and N (blue) atoms,
but for clarity the TFSI anions beyond interfaces are represented by isosurfaces. 
(b) Density profiles for Li (blue), water (orange) and TFSI (green) at $\Delta\Psi=0$~V (top) and 3~V (bottom).
(c) Density profiles for O and F from TFSI anions on the positive electrode, for $\Delta\Psi=1.5$ and 2~V; the shaded areas indicate the first and second adsorbed layers; the key indicates the potential of the positive electrode, assumed to be $\Psi^+=\Delta\Psi/2$). 
(d,e) Snapshots illustrating the structure close to the positive electrode from simulations at $\Delta\Psi=1.5$ and 2~V.
Adapted with permission from Ref.~\cite{li2018b}, \emph{J. Phys. Chem. C} 2018, {\bf 122}, 23917, Copyright (2018) American Chemical Society.
}
\label{fig:structure}
\end{figure}

As an illustration, we consider here the complex case of a water-in-salt electrolyte (WiSE), proposed recently as promising for high-voltage batteries and supercapacitors~\cite{suo2013a,wang2016j}, by reporting some results of Li \emph{et al.}~\cite{li2018b}. Figure~\ref{fig:structure}a shows a system consisting of a 20~mol/kg aqueous lithium bis-(trifluoromethane)sulfonimide (LiTFSI) solution between two graphite electrodes maintained at a constant potential using the method described in section~\ref{sec:fluctuatingcharges}. In such an electrolyte, the number of water molecules per ion pair is lower than 3 (compared to $\approx$50 for a typical 1~mol/L concentration), so that standard EDL theories are not expected to hold. The density profiles in Figure~\ref{fig:structure}b clearly illustrate the layering of the interfacial fluid even in the absence of applied voltage and a bulk region far from the surfaces, as well as the change in composition and structure of both interfaces under voltage. One can note in particular the increase in water concentration near the negative electrode (left) accompanying that of Li$^+$, which approach the surface with their limited hydration shell. The structure on the positive electrode is further illustrated in panel~\ref{fig:structure}c, which shows the density profiles for oxygen and fluorine atoms from the TFSI anion for two voltages. The most dramatic structural change occurring at this surface between $\Delta\Psi=1.5$ and 2~V is the reorientation of the anions, leading to a closer approach of O atoms (carrying a larger partial charge than F) to the surface, also visible on the typical snapshots of panels ~\ref{fig:structure}d and~\ref{fig:structure}e. Even though we do not discuss this further here, it was also found in this work, using importance sampling techniques, that changes in the interfacial composition and structure as a function of voltages could be linked to peaks in the differential capacitance~\cite{li2018b}, as previously reported for room temperature ionic liquids~\cite{merlet2014a,rotenberg2015a} and concentrated electrolytes~\cite{uralcan2016a}. One can finally note that, even in such complex cases, molecular simulation results can be used as reference data to validate and parameterize simpler theories, as was done \emph{e.g.} by McEldrew \emph{et al.} in a contemporary study of a similar WiSE between uniformly charged walls~\cite{mceldrew2018a}.

\subsection{Interfacial dynamics}
\label{sec:dynamics}

Applications of electrode-electrolyte interfaces to energy storage (batteries, supercapacitors) not only aim at improving the amount of energy that can be stored, by increasing the capacitance of the interface or the voltage, but also at reducing the charge/discharge time to increase the power that can be delivered. In the standard $RC$ circuit picture, the charging time depends on the resistance of the bulk electrolyte and the interfacial capacitance, but not on the interfacial dynamics. The situation in realistic porous electrode materials is more complex and molecular simulation has contributed to a better understanding of the charging dynamics in these systems, also making the link with more involved equivalent circuit models used in electrochemistry~\cite{pean2014a,pean2015b,pean2016b,breitsprecher2018a}. However, even in the simpler case of a planar electrode and a solvent-based electrolyte, which is the focus of the present work, the dynamics of the interfacial fluid is modified by the presence of the solid. This in turn may modify not only the charging dynamics but also the kinetics of electron transfer reactions (which will be discussed in more detail in section~\ref{sec:electrochemistry}). We illustrate here some important findings on the dynamics at a water-platinum interface uncovered using molecular simulation.

\begin{figure}[ht!]
\includegraphics[width=5in]{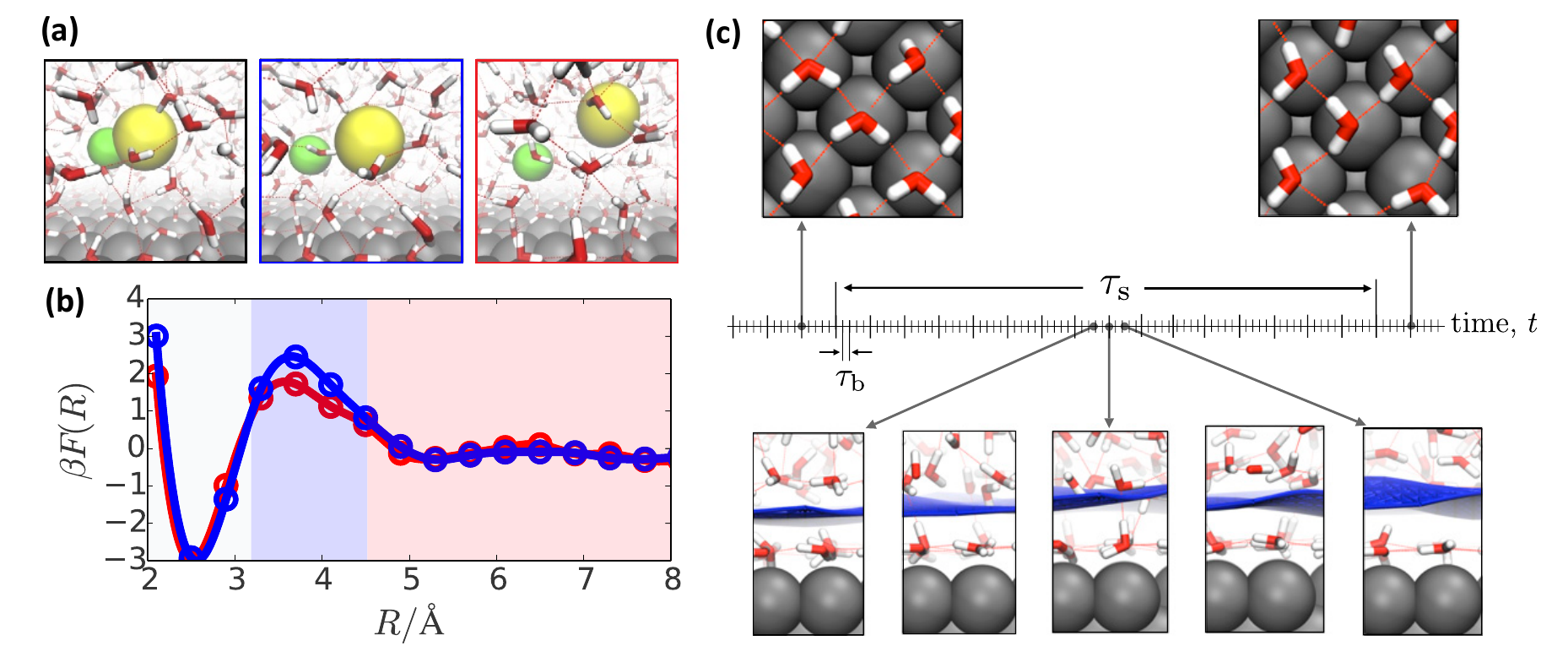}
\caption{
(a) Typical configurations for the dissociation of an ion pair (Na$^+$ in green, I$^-$ in yellow) in water near a (111) Pt electrode surface, from a contact ion pair (black-bordered panel) to a dissociated state (red-bordered panel).
(b) Free energy profile as a function (in units of the thermal energy $k_BT=1/\beta$) of the distance $R$ between the ions in the bulk (red) and near the surface (blue).
(c) Illustration of the slow time scale $\tau_s$ associated with structural rearrangements of water near a (100) Pt surface, compared to that $\tau_b$ for the relaxation of bulk density fluctuations (small ticks are separated by 20~ps, larger than typical values for 
$\tau_b$). The blue surface in the bottom panels indicates the instantaneous interface between the first adsorbed layer and the rest of the fluid, which fluctuates due to local changes in the hydrogen bond pattern. Collective fluctuations eventually leading to global rearrangements such as between the top panels occur on much longer time scales.
Panels (a) and (b) adapted with permission from Ref.~\cite{kattirtzi2017a}, \emph{PNAS}, 2017, {\bf 114}, 13374; panel (c) adapted with permission from Ref.~\cite{limmer2013b}, with permission of \emph{PNAS}, 2013, {\bf 110}, 4200.
}
\label{fig:dynamics}
\end{figure}

Kattirtzi \emph{et al.} investigated the dissociation of ion pairs near a Pt electrode using classical (for Na$^+$-I$^-$) and \emph{ab initio} (for water ions H$_3$O$^+$-HO$^-$) molecular dynamics~\cite{kattirtzi2017a}. Figure~\ref{fig:dynamics}a shows configurations for the Na$^+$-I$^-$ case near a (111) Pt electrode surface, modeled using the method described in section~\ref{sec:fluctuatingcharges} and the force field of Siepmann and Sprik~\cite{siepmann1995a} for the water-Pt interaction, which includes two and three body terms. The three configurations are typical for a contact ion pair, an intermediate state and a dissociated state, corresponding to the regions visible in the free energy profiles as a function of the interionic distance $R$, obtained by umbrella sampling (see section~\ref{sec:importancesampling}) and shown in panel~\ref{fig:dynamics}b. This panel further shows that the free energy profile along this distance $R$ is in fact only slightly affected by the vicinity of the surface, with no change in the relative free energies of the states (suggesting limited changes in their solvation structure with respect to the bulk) and a small increase in the activation barrier for the dissociation/recombination, which would suggest only a slight decrease in the corresponding rates. However, as in the bulk~\cite{mullen2014a}, the interionic distance is not sufficient to examine the mechanism. It also involves collective water fluctuations, which were described by Kattirtzi \emph{et al.} using a simple collective variable, namely the Madelung potential on the ions. From the change in free energy barrier including this variable (with an increase of 1.5~$k_BT$ with respect to the bulk), one would predict a decrease in the dissociation rate near the electrode $k_{elec}\approx0.2~k_{bulk}$. However the dissociation rate measured in the simulations is in fact $k_{elec}\approx0.02~k_{bulk}$, \emph{i.e.} the decrease is ten times larger. This is because the free energy is not sufficient to predict the rate, as the dynamics of water fluctuations also plays a role~\cite{kattirtzi2017a}.

The dynamics of water near a Pt surface was in fact considered in earlier work by some of the same authors~\cite{limmer2013b,willard2013a,limmer2015b,limmer2015c}. It was found that these dynamics can be highly collective, heterogeneous and slow. Figure~\ref{fig:dynamics}c illustrates the slow time scales associated with structural rearrangements of water near a (100) Pt surface, compared to that for the relaxation of bulk density fluctuations. On this particular surface, water molecules in the first adsorbed layer form an ordered network of strong H-bonds between them. The lack of interactions with molecules beyond this layer (see the bottom parts of this panel) in fact results in an unexpected hydrophobic behavior, with \emph{e.g.} a large contact angle of a water droplet on this adlayer (see section~\ref{sec:electrowetting} for more discussion of contact angles). From the dynamical point of view, switching between distinct but equally probable H-bond patterns (see top parts of Figure~\ref{fig:dynamics}c) requires collective rearrangements, associated with very long time scales and displaying dynamical heterogeneity~\cite{limmer2013b}. While some results observed on this particular surface should apply generally to other Pt surfaces and other metals, the details of the dynamics crucially depend on the interplay between the geometry of the solid lattice and the interactions with the fluid, so that molecular simulation is ideally suited for such investigations. It was for example used recently to analyze the reorientation dynamics of water molecules at an electrified graphene interface~\cite{zhang2020a}.

\subsection{Electrowetting}
\label{sec:electrowetting}

\begin{figure}[ht!]
\includegraphics[width=4.5in]{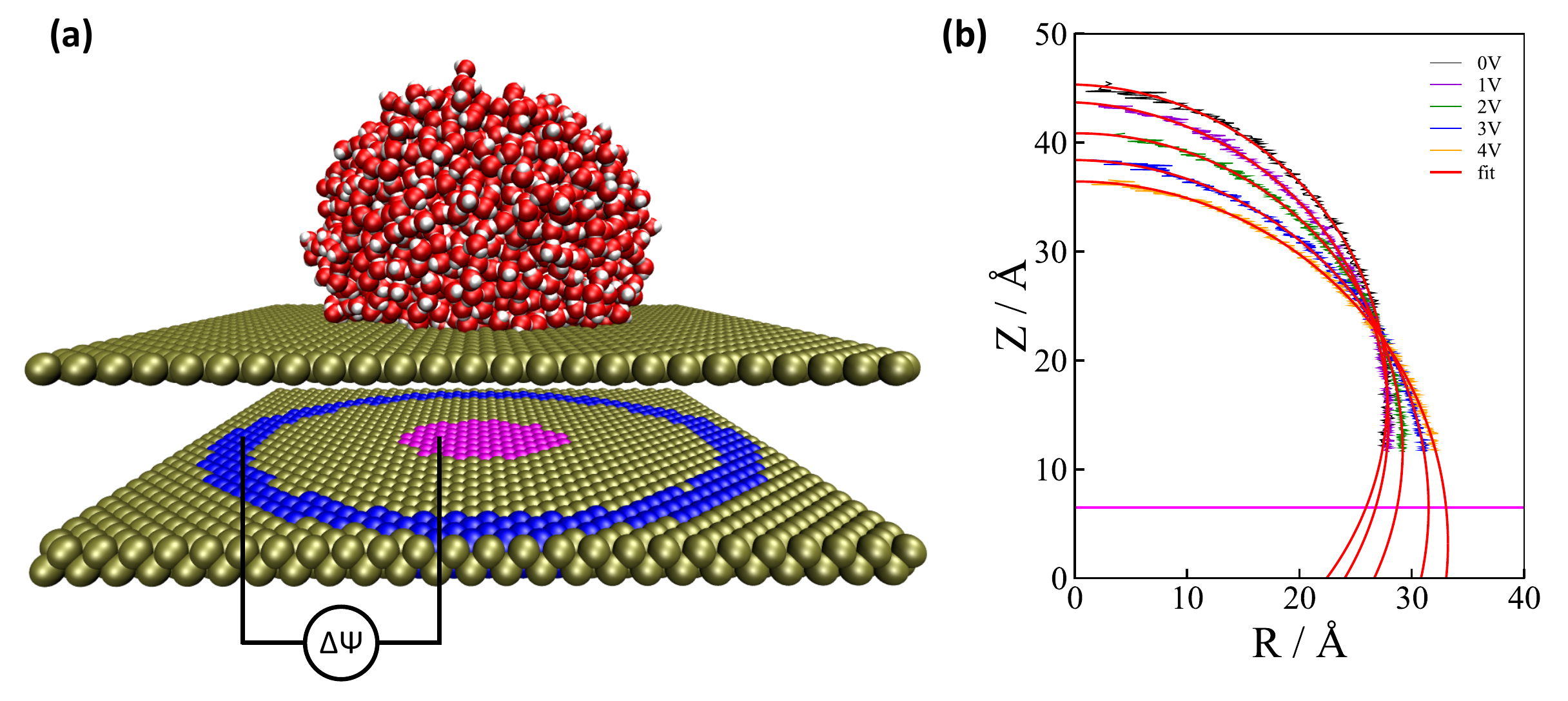}
\caption{
(a) Representation of the electrowetting-on-dielectric simulated system. Ochre atoms below the aqueous drop constitute a hydrophobic insulator monolayer. The magenta and blue atoms are the inner and outer platinum electrodes, which are also separated by an insulator ring (a vertical shift was introduced between the two layers of the substrate material for a better visualization).
(b) Vertical cross section of the water droplet on the insulator surface at various applied voltages. The contact angle is determined by a fit of the drop contour.
Adapted with permission from Ref.~\cite{choudhuri2016a}, \emph{ACS Nano} 2016, {\bf 10}, 8536, Copyright (2016) American Chemical Society.
}
\label{fig:electrowetting}
\end{figure}

Electrochemistry also goes beyond simple setups with two electrodes facing each other and an electrolyte in between. For example, electrowetting-on-dielectric setups were proposed for switchable optical devices. An example of design is made of two concentric electrodes (with different radii) separated from the liquid, which is made of a water nanodrop, by a layer of dielectric~\cite{krupenkin2003a}. The use of constant potential simulations allowed an accurate representation of such systems at the molecular scale~\cite{choudhuri2016a}, as shown on Figure \ref{fig:electrowetting}a. Applying a potential between the two electrodes results in a noticeable spreading of the drop, an effect which is reversible. Figure  \ref{fig:electrowetting}b shows the cross-section profiles at various applied potentials, from which contact angles can be extracted. In this example the angle varies from 111$^\circ$ at null voltage down to 84~$^\circ$ at 4~V. Changing the radii of the two electrodes leads to different values but the spreading effect persists. At the macroscale, the variation of the contact angle with the applied potential is well predicted by the Young-Lippmann equation, but simulations showed that it does not hold anymore at the nanometric scales. By performing additional non-equilibrium simulations where the potential was suddenly applied or released, it was observed that the retraction at zero voltage is much faster than the corresponding spreading, an effect which could be explained by different liquid/solid friction properties.

\subsection{Towards electrochemistry}
\label{sec:electrochemistry}

In general, simulating reactive systems is challenging in molecular simulation: Reactive force fields need to be employed, which are difficult to parameterize. Nevertheless, many electrochemical reactions do not involve bond breaking/formation, but only a change of the redox state of the species. This observation is central in Marcus theory, which was developed in order to calculate reaction free energies for electron transfer~\cite{marcus1956a}. In short, this classical transition state theory introduces a key quantity, the solvent reorganization energy, associated with the solvent rearrangement when the redox state of the solute changes. In classical molecular simulation, Warshel introduced the vertical energy gap, that is the potential energy difference between the oxidized and the reduced species \emph{at fixed solvent coordinates}, as the key reaction coordinate for simulating electrochemical reactions~\cite{warshel1982a}.

Although this development allowed straightforward studies  of electron transfer reactions in bulk solutions, in an electrochemical cell setup an additional complexity arises because the electrons are transferred to the electrodes instead of vacuum. Madden and co-workers proposed an approach in which the constant applied potential condition is enforced before and after the electron transfer reactions, leading to different charge organizations inside the electrodes~\cite{reed2008a}. Using this approach, they showed how the solvent reorganization energy, and hence the free energy curves, depend on the applied potential and on the distance of the redox species from the electrode in a molten salt. Surprisingly, the variations they observed where much smoother than expected from the pronounced oscillations in the mean electrical potential inside the double-layer region, a result which was confirmed in further studies~\cite{pounds2015a}.

The next challenge for classical simulations relies on the study of active interface. First steps in this direction were made using the Empirical Valence Bond method to describe the transfer of proton in water on silver or platinum electrodes~\cite{wilhelm_modeling_2011, wiebe_water_2017}. Switching in electrometallization cells has also been studied using reactive force fields (REAXFF) together with the charge equilibration method for the electrostatic part~\cite{onofrio2015a,onofrio2015b}. The electrodes, made of copper, were allowed to dissolve inside an amorphous silica and redeposit upon application/cancellation of a voltage, eventually leading to the formation of a conductive contact between the two electrodes. The extension of this approach to more conventional solvents is yet hindered by the difficulty of developing accurate reactive force fields for complex systems. Another strategy was recently proposed, in which the charge transfer is modeled as a stochastic process, using methods borrowed from grand canonical Monte Carlo simulations~\cite{dwelle2019a}. We can therefore expect that more simulations of reactive electrochemical interfaces will be proposed in future years.

\section{SUMMARY POINTS}
\begin{enumerate}
\item Over the last 30 years, classical molecular simulations have emerged as an essential tool to investigate the properties of electrode-electrolyte interfaces: By providing a compromise between an atomic description and a computational cost allowing a sufficient sampling of relevant electrolyte configurations, they offer the possibility to face the challenge posed by such interfaces bridging electrons in a solid and ions in a solvent, \emph{i.e.} Quantum Chemistry and Statistical Physics.
\item We have provided an overview of the models and methods to describe the metallic character of the electrode and its interactions with the electrolyte solution, and we discussed simulation setups, algorithms and statistical tools to sample configurations from the statistical ensemble corresponding to constant-potential electrodes.
\item We have illustrated a selection of properties which can be investigated with classical molecular simulations, with examples on the capacitance, the interfacial structure and dynamics, electrowetting, as well as steps towards electrochemistry. 
\item The results obtained at this classical level can be compared directly to experiments probing the interface on the molecular scale, but also serve as reference data for simpler theories of electric double layers, generally used to interpret macroscopic electrochemical experiments.
\item Even though we restricted ourselves to the simpler yet practically relevant and physically rich case of planar interfaces and solvent-based liquids, from pure solvent to water-in-salt-electrolytes, some of the methods discussed here also apply to more complex (\emph{e.g.} disordered nanoporous) electrodes and to solvent-free electrolytes (room temperature ionic liquids).
\end{enumerate}

\section{FUTURE ISSUES}
\begin{enumerate}
\item The classical description of a metallic electrode remains a challenging issue. For example, most electrode materials do not behave as perfect metals. Electrostatic screening inside the solid can be captured at the continuum level using Thomas-Fermi theory, an approach which can also be exploited for classical molecular simulation. Earlier attempts included this effect of screening on the electrolyte via an external potential~\cite{rose_solvation_1991,calhoun_electron_1996}, but new strategies have been proposed very recently, based on fluctuating charges~\cite{TFarXiv} or on mobile charges~\cite{schlaich2020arXiv}. This will be particularly useful to address the effect of the metallic character of the electrode on the properties of the interfacial fluids, such as the nanoscale capillary freezing of ionic liquids~\cite{comtet2017a}.
\item The treatment of non-electrostatic interactions also offers opportunities for improvement. The flexibility of the electrode has almost never been addressed, even though it might play a role in particular in porous electrodes. Due to the importance of electrochemical reactions, it is likely that reactive force fields will also gain importance, if their parametrization can be extended to more complex systems. The availability of accurate reference data from \emph{ab initio} calculations, in particular quantum Monte Carlo, will allow a better parametrization of force fields based on existing models. One can also anticipate that the rapid development of Machine-Learning based approaches (\emph{e.g.} force fields based on neural networks) in molecular simulation will also reach the community working on electrode-electrolyte interfaces. 
\item Another promising strategy is offered by hybrid approaches, coupling several levels of descriptions. As an example, a mesoscopic description of the solvent, based on classical density functional theory, was recently coupled to a fluctuating charge model of a graphene electrode~\cite{jeanmairet2019b} and used to investigate electron transfer reactions in a classical solute~\cite{jeanmairet2019a}. In the other direction, the development of classical models of the interface might also help improving the coupling between the electrolyte and the electrode in QM/MM simulations.
\item From the physical point of view, all these developments should allow to investigate new phenomena such as the electro-mechanical couplings related to electrotunable lubricity~\cite{fajardo2015a} or Electrochemical Quartz Microbalance Experiments. The ability to take bond formation an breaking into account would also open the way to the study of the formation of the so-called solid electrolyte interphase (SEI), which plays a crucial role in batteries.
\end{enumerate}

\section*{DISCLOSURE STATEMENT}
The authors are not aware of any affiliations, memberships, funding, or financial holdings that
might be perceived as affecting the objectivity of this review.

\section*{ACKNOWLEDGMENTS}
The authors are grateful to many colleagues with whom they have worked or exchanged on this topic over the years, in particular Paul Madden, Michiel Sprik, David Limmer, C\'eline Merlet. 
This project has received funding from the European Research Council (ERC) under the European Union's Horizon 2020 research and innovation programme (grant agreement No. 771294 and grant agreement No. 863473). This work was supported by the French National Research Agency (Labex STORE-EX, Grant  ANR-10-LABX-0076, and project NEPTUNE, Grant ANR-17-CE09-0046-02).

%



\begin{thebibliography}{100}

\bibitem{salanne2016a}
M.~Salanne, B.~Rotenberg, K.~Naoi, K.~Kaneko, P.-L. Taberna, C.~P. Grey,
  B.~Dunn, and P.~Simon.
\newblock {Efficient Storage Mechanisms for Building Better Supercapacitors}.
\newblock {\em Nat. Energy}, 1:16070, 2016.

\bibitem{seh_combining_2017}
Zhi~Wei Seh, Jakob Kibsgaard, Colin~F. Dickens, Ib~Chorkendorff, Jens~K.
  Norskov, and Thomas~F. Jaramillo.
\newblock Combining theory and experiment in electrocatalysis: {Insights} into
  materials design.
\newblock {\em Science}, 355(6321), January 2017.

\bibitem{parsons1990a}
R.~Parsons.
\newblock The electrical double layer: recent experimental and theoretical
  developments.
\newblock {\em Chem. Rev.}, 90:813--826, 1990.

\bibitem{gouy1910a}
G.~Gouy.
\newblock Sur la constitution de la charge \'electrique \`a la surface d'un
  \'electrolyte.
\newblock {\em J. Phys. Theor. Appl}, 9:457, 1910.

\bibitem{chapman1913a}
D.~Chapman.
\newblock {LI. A contribution to the theory of electrocapillarity}.
\newblock {\em Phil. Mag.}, 25:475, 1913.

\bibitem{stern1924a}
O.~Stern.
\newblock {Zur Theorie Der Elektrolytischen Doppelschicht}.
\newblock {\em Z. Elektrochem.}, 30:508, 1924.

\bibitem{bazant2011a}
M.~Z. Bazant, B.~D. Storey, and A.~A. Kornyshev.
\newblock Double layer in ionic liquids: Overscreening versus crowding.
\newblock {\em Phys. Rev. Lett.}, 106:046102, 2011.

\bibitem{goodwin2017a}
Z.~A.~H. Goodwin, G.~Feng, and A.~A. Kornyshev.
\newblock Mean-field theory of electrical double layer in ionic liquids with
  account of short-range correlations.
\newblock {\em Electrochim. Acta}, 225:190--197, 2017.

\bibitem{mceldrew2018a}
Michael McEldrew, Zachary A.~H. Goodwin, Alexei~A. Kornyshev, and Martin~Z.
  Bazant.
\newblock Theory of the {Double} {Layer} in {Water}-in-{Salt} {Electrolytes}.
\newblock {\em The Journal of Physical Chemistry Letters}, 9(19):5840--5846,
  2018.

\bibitem{bazant2004a}
Martin~Z. Bazant, Katsuyo Thornton, and Armand Ajdari.
\newblock Diffuse-charge dynamics in electrochemical systems.
\newblock {\em Physical Review E}, 70(2):021506, August 2004.

\bibitem{janssen2018a}
Mathijs Janssen and Markus Bier.
\newblock Transient dynamics of electric double layer capacitors: {Exact}
  expressions within the {Debye}-{Falkenhagen} approximation.
\newblock {\em {P}hysical {R}eview {E}}, 97:052616, February 2018.

\bibitem{netz2003a}
R.~R. Netz.
\newblock Electrofriction and {Dynamic} {Stern} {Layers} at {Planar} {Charged}
  {Surfaces}.
\newblock {\em Physical Review Letters}, 91(13):138101, 2003.

\bibitem{grun2004a}
F.~Gr\"un, M.~Jardat, P.~Turq, and C.~Amatore.
\newblock Relaxation of the electrical double layer after an electron transfer
  approached by \uppercase{B}rownian dynamics simulation.
\newblock {\em J. Chem. Phys.}, 120(20):9648, 2004.

\bibitem{pagonabarraga2010a}
I.~Pagonabarraga, B.~Rotenberg, and D.~Frenkel.
\newblock Recent advances in the modelling and simulation of electrokinetic
  effects: bridging the gap between atomistic and macroscopic descriptions.
\newblock {\em Phys. Chem. Chem. Phys.}, 12(33):9566--9580, 2010.

\bibitem{lobaskin2016a}
Vladimir Lobaskin and Roland~R. Netz.
\newblock Diffusive-convective transition in the non-equilibrium charging of an
  electric double layer.
\newblock {\em EPL (Europhysics Letters)}, 116(5):58001, 2016.

\bibitem{asta2019a}
Adelchi~J. Asta, Ivan Palaia, Emmanuel Trizac, Maximilien Levesque, and
  Benjamin Rotenberg.
\newblock Lattice {Boltzmann} electrokinetics simulation of nanocapacitors.
\newblock {\em The Journal of Chemical Physics}, 151(11):114104, September
  2019.

\bibitem{striolo2016a}
A.~Striolo, A.~Michaelides, and L.~Joly.
\newblock The carbon-water interface: Modeling challenges and opportunities for
  the water-energy nexus.
\newblock {\em Annu. Rev. Chem. Biomol. Eng.}, 7:533--556, 2016.

\bibitem{brandenburg2019a}
Jan~Gerit Brandenburg, Andrea Zen, Martin Fitzner, Benjamin Ramberger, Georg
  Kresse, Theodoros Tsatsoulis, Andreas Gr\"uneis, Angelos Michaelides, and
  Dario Alf\`e.
\newblock Physisorption of {Water} on {Graphene}: {Subchemical} {Accuracy} from
  {Many}-{Body} {Electronic} {Structure} {Methods}.
\newblock {\em The Journal of Physical Chemistry Letters}, 10(3):358--368,
  2019.

\bibitem{taylor2006a}
Christopher~D. Taylor, Sally~A. Wasileski, Jean-Sebastien Filhol, and Matthew
  Neurock.
\newblock First principles reaction modeling of the electrochemical interface:
  {Consideration} and calculation of a tunable surface potential from atomic
  and electronic structure.
\newblock {\em Physical Review B}, 73(16):165402, 2006.

\bibitem{lautar2020a}
Anja~Kopa\v{c} Lautar, Arthur Hagopian, and Jean-S\'ebastien Filhol.
\newblock Modeling interfacial electrochemistry: concepts and tools.
\newblock {\em Physical Chemistry Chemical Physics}, 22(19):10569--10580, 2020.
\newblock Publisher: The Royal Society of Chemistry.

\bibitem{merlet2012a}
C.~Merlet, B.~Rotenberg, P.~A. Madden, P.-L. Taberna, P.~Simon, Y.~Gogotsi, and
  M.~Salanne.
\newblock {On the Molecular Origin of Supercapacitance in Nanoporous Carbon
  Electrodes}.
\newblock {\em Nat. Mater.}, 11:306--310, 2012.

\bibitem{merlet2013c}
C.~Merlet, B.~Rotenberg, P.~A. Madden, and M.~Salanne.
\newblock Computer simulations of ionic liquids at electrochemical interfaces.
\newblock {\em Phys. Chem. Chem. Phys.}, 15:15781--15792, 2013.

\bibitem{simoncelli2018a}
M.~Simoncelli, N.~Ganfoud, A.~Sene, M.~Haefele, B.~Daffos, P.-L. Taberna,
  M.~Salanne, P.~Simon, and B.~Rotenberg.
\newblock Blue energy and desalination with nanoporous carbon electrodes:
  Capacitance from molecular simulations to continuous models.
\newblock {\em Phys. Rev. X}, 8:021024, 2018.

\bibitem{merlet2014a}
C.~Merlet, D.~T. Limmer, M.~Salanne, R.~{van Roij}, P.~A. Madden, D.~Chandler,
  and B.~Rotenberg.
\newblock The electric double layer has a life of its own.
\newblock {\em J. Phys. Chem. C}, 118:18291--18298, 2014.

\bibitem{fedorov2014a}
M.~V. Fedorov and A.~A. Kornyshev.
\newblock Ionic liquids at electrified interfaces.
\newblock {\em Chem. Rev.}, 114:2978---3036, 2014.

\bibitem{burt2014a}
R.~Burt, G.~Birkett, and X.~S. Zhao.
\newblock A review of molecular modelling of electric double layer capacitors.
\newblock {\em Phys. Chem. Chem. Phys.}, 16:6519--6538, 2014.

\bibitem{burt2016a}
R.~Burt, K.~Breitsprecher, B.~Daffos, P.-L. Taberna, P.~Simon, G.~Birkett,
  X.~S. Zhao, C.~Holm, and M.~Salanne.
\newblock Capacitance of nanoporous carbon-based supercapacitors is a trade-off
  between the concentration and the separability of the ions.
\newblock {\em J. Phys. Chem. Lett.}, 7:4015--4021, 2016.

\bibitem{li2017j}
Z.~Li, G.~Jeanmairet, T.~{Mendez-Morales}, M.~Burbano, M.~Haefele, and
  M.~Salanne.
\newblock Confinement effects on an electron transfer reaction in nanoporous
  carbon electrodes.
\newblock {\em J. Phys. Chem. Lett.}, 8:1925--1931, 2017.

\bibitem{torrie_electrical_1980}
G.~M. Torrie and J.~P. Valleau.
\newblock Electrical double layers. {I}. {Monte} {Carlo} study of a uniformly
  charged surface.
\newblock {\em The Journal of Chemical Physics}, 73(11):5807--5816, December
  1980.

\bibitem{glosli_molecular_1992}
James~N. Glosli and Michael~R. Philpott.
\newblock Molecular dynamics simulation of adsorption of ions from aqueous
  media onto charged electrodes.
\newblock {\em The Journal of Chemical Physics}, 96(9):6962--6969, May 1992.

\bibitem{kiyohara_monte_2007}
Kenji Kiyohara and Kinji Asaka.
\newblock Monte {Carlo} simulation of electrolytes in the constant voltage
  ensemble.
\newblock {\em The Journal of Chemical Physics}, 126(21):214704, June 2007.

\bibitem{kiyohara_monte_2007-1}
Kenji Kiyohara and Kinji Asaka.
\newblock Monte {Carlo} {Simulation} of {Porous} {Electrodes} in the {Constant}
  {Voltage} {Ensemble}.
\newblock {\em The Journal of Physical Chemistry C}, 111(43):15903--15909,
  November 2007.

\bibitem{van_megen_grand_1980}
William van Megen and Ian Snook.
\newblock The grand canonical ensemble {Monte} {Carlo} method applied to the
  electrical double layer.
\newblock {\em The Journal of Chemical Physics}, 73(9):4656--4662, November
  1980.

\bibitem{crozier_molecular-dynamics_2001}
Paul~S. Crozier, Richard~L. Rowley, and Douglas Henderson.
\newblock Molecular-dynamics simulations of ion size effects on the fluid
  structure of aqueous electrolyte systems between charged model electrodes.
\newblock {\em The Journal of Chemical Physics}, 114(17):7513--7517, May 2001.

\bibitem{lee_molecular_1986}
Song~Hi Lee, Jayendran~C. Rasaiah, and J.~B. Hubbard.
\newblock Molecular dynamics study of a dipolar fluid between charged plates.
\newblock {\em The Journal of Chemical Physics}, 85(9):5232--5237, November
  1986.
\newblock Publisher: American Institute of Physics.

\bibitem{nagy_molecular_1990}
G.~Nagy and K.~Heinzinger.
\newblock A {Molecular} {Dynamics} simulation of electrified platinum/water
  interfaces.
\newblock {\em Journal of Electroanalytical Chemistry and Interfacial
  Electrochemistry}, 296(2):549--558, December 1990.

\bibitem{hautman_molecular_1989}
J.~Hautman, J.~W. Halley, and Y.-J. Rhee.
\newblock Molecular dynamics simulation of water beween two ideal classical
  metal walls.
\newblock {\em The Journal of Chemical Physics}, 91(1):467--472, July 1989.
\newblock Publisher: American Institute of Physics.

\bibitem{watanabe_dielectric_1991}
Masakatsu Watanabe, Anatol~M. Brodsky, and William~P. Reinhardt.
\newblock Dielectric properties and phase transitions of water between
  conducting plates.
\newblock {\em The Journal of Physical Chemistry}, 95(12):4593--4596, June
  1991.

\bibitem{rose_adsorption_1993}
Daniel~A. Rose and Ilan Benjamin.
\newblock Adsorption of {Na}$^+$ and {Cl}$^-$ at the charged water--platinum
  interface.
\newblock {\em The Journal of Chemical Physics}, 98(3):2283--2290, February
  1993.

\bibitem{smith_simulation_1994}
B.~B. Smith and J.~W. Halley.
\newblock Simulation study of the ferrous ferric electron transfer at a
  metal--aqueous electrolyte interface.
\newblock {\em The Journal of Chemical Physics}, 101(12):10915--10924, December
  1994.

\bibitem{zhu_structure_1991}
S.-B. Zhu and G.~W. Robinson.
\newblock Structure and dynamics of liquid water between plates.
\newblock {\em The Journal of Chemical Physics}, 94(2):1403--1410, January
  1991.
\newblock Publisher: American Institute of Physics.

\bibitem{daub_electrowetting_2007}
Christopher~D. Daub, Dusan Bratko, Kevin Leung, and Alenka Luzar.
\newblock Electrowetting at the {Nanoscale}.
\newblock {\em The Journal of Physical Chemistry C}, 111(2):505--509, January
  2007.

\bibitem{ilja_siepmann_ordering_1992}
J.~Ilja~Siepmann and Michiel Sprik.
\newblock Ordering of fractional monolayers of {H$_2$O} on {Ni}(110).
\newblock {\em Surface Science Letters}, 279(1):L185--L190, December 1992.

\bibitem{siepmann1995a}
J.~I. Siepmann and M.~Sprik.
\newblock {Influence of Surface-Topology and Electrostatic Potential on Water
  Electrode Systems}.
\newblock {\em J. Chem. Phys.}, 102:511--524, 1995.

\bibitem{guymon_simulating_2005}
{Guymon}, {Rowley}, {Harb}, and {Wheeler}.
\newblock Simulating an electrochemical interface using charge dynamics.
\newblock {\em Condensed Matter Physics}, 8(2):335, 2005.

\bibitem{reed2007a}
S.~K. Reed, O.~J. Lanning, and P.~A. Madden.
\newblock {Electrochemical Interface Between an Ionic Liquid and a Model
  Metallic Electrode}.
\newblock {\em J. Chem. Phys.}, 126:084704, 2007.

\bibitem{pounds2009b}
M.~Pounds, S.~Tazi, M.~Salanne, and P.~A. Madden.
\newblock Ion adsorption at a metallic electrode: an {\it ab initio} based
  simulation study.
\newblock {\em J. Phys.: Condens. Matter}, 21:424109, 2009.

\bibitem{vatamanu_molecular_2009}
Jenel Vatamanu, Oleg Borodin, and Grant~D. Smith.
\newblock Molecular dynamics simulations of atomically flat and nanoporous
  electrodes with a molten salt electrolyte.
\newblock {\em Physical Chemistry Chemical Physics}, 12(1):170--182, December
  2009.
\newblock Publisher: The Royal Society of Chemistry.

\bibitem{petersen2012a}
M.~K. Petersen, R.~Kumar, H.~S. White, and G.~A. Voth.
\newblock A computationally efficient treatment of polarizable electrochemical
  cells held at a constant potential.
\newblock {\em J. Phys. Chem. C}, 116:4903--4912, 2012.

\bibitem{torrie_electrical_1982}
G.~M. Torrie, J.~P. Valleau, and G.~N. Patey.
\newblock Electrical double layers. {II}. {Monte} {Carlo} and {HNC} studies of
  image effects.
\newblock {\em The Journal of Chemical Physics}, 76(9):4615--4622, May 1982.

\bibitem{parsonage_computer_1986}
Neville~G. Parsonage and David Nicholson.
\newblock Computer simulation of water between metal walls.
\newblock {\em Journal of the Chemical Society, Faraday Transactions 2:
  Molecular and Chemical Physics}, 82(9):1521--1535, January 1986.
\newblock Publisher: The Royal Society of Chemistry.

\bibitem{gardner_waterlike_1987}
Anne~A. Gardner and John~P. Valleau.
\newblock Water--like particles at surfaces. {II}. {In} a double layer and at a
  metallic surface.
\newblock {\em The Journal of Chemical Physics}, 86(7):4171--4176, April 1987.
\newblock Publisher: American Institute of Physics.

\bibitem{klapp_monte-carlo_2006}
Sabine H.~L. Klapp.
\newblock Monte-{Carlo} simulations of strongly interacting dipolar fluids
  between two conducting walls.
\newblock {\em Molecular Simulation}, 32(8):609--621, July 2006.

\bibitem{takae_fluctuations_2015}
Kyohei Takae and Akira Onuki.
\newblock Fluctuations of local electric field and dipole moments in water
  between metal walls.
\newblock {\em The Journal of Chemical Physics}, 143(15):154503, October 2015.

\bibitem{girotto_simulations_2017}
Matheus Girotto, Alexandre~P. dos Santos, and Yan Levin.
\newblock Simulations of ionic liquids confined by metal electrodes using
  periodic {Green} functions.
\newblock {\em The Journal of Chemical Physics}, 147(7):074109, August 2017.

\bibitem{tyagi_icmmm2d_2007}
Sandeep Tyagi, Axel Arnold, and Christian Holm.
\newblock {ICMMM2D}: {An} accurate method to include planar dielectric
  interfaces via image charge summation.
\newblock {\em The Journal of Chemical Physics}, 127(15):154723, October 2007.
\newblock Publisher: American Institute of Physics.

\bibitem{arnold2013a}
A.~Arnold, K.~Breitsprecher, F.~Fahrenberger, S.~Kesselheim, O.~Lenz, and
  C.~Holm.
\newblock Efficient algorithms for electrostatic interactions including
  dielectric contrasts.
\newblock {\em Entropy}, 15:4569--4588, 2013.

\bibitem{allen_electrostatic_2001}
Rosalind Allen, Jean-Pierre Hansen, and Simone Melchionna.
\newblock Electrostatic potential inside ionic solutions confined by
  dielectrics: a variational approach.
\newblock {\em Physical Chemistry Chemical Physics}, 3(19):4177--4186, January
  2001.
\newblock Publisher: The Royal Society of Chemistry.

\bibitem{boda_computing_2004}
Dezsö Boda, Dirk Gillespie, Wolfgang Nonner, Douglas Henderson, and Bob
  Eisenberg.
\newblock Computing induced charges in inhomogeneous dielectric media:
  {Application} in a {Monte} {Carlo} simulation of complex ionic systems.
\newblock {\em Physical Review E}, 69(4):046702, April 2004.
\newblock Publisher: American Physical Society.

\bibitem{tyagi_iterative_2010}
Sandeep Tyagi, Mehmet Süzen, Marcello Sega, Marcia Barbosa, Sofia~S.
  Kantorovich, and Christian Holm.
\newblock An iterative, fast, linear-scaling method for computing induced
  charges on arbitrary dielectric boundaries.
\newblock {\em The Journal of Chemical Physics}, 132(15):154112, April 2010.
\newblock Publisher: American Institute of Physics.

\bibitem{breitsprecher2015a}
K.~Breitsprecher, K.~Szuttor, and C.~Holm.
\newblock Electrode models for ionic liquid-based capacitors.
\newblock {\em J. Phys. Chem. C}, 119:22445--22451, 2015.

\bibitem{barros_efficient_2014}
Kipton Barros, Daniel Sinkovits, and Erik Luijten.
\newblock Efficient and accurate simulation of dynamic dielectric objects.
\newblock {\em The Journal of Chemical Physics}, 140(6):064903, February 2014.

\bibitem{geada_insight_2018}
Isidro~Lorenzo Geada, Hadi Ramezani-Dakhel, Tariq Jamil, Marialore Sulpizi, and
  Hendrik Heinz.
\newblock Insight into induced charges at metal surfaces and biointerfaces
  using a polarizable {Lennard}--{Jones} potential.
\newblock {\em Nature Communications}, 9(1):716, December 2018.

\bibitem{iori_including_2008}
F.~Iori and S.~Corni.
\newblock Including image charge effects in the molecular dynamics simulations
  of molecules on metal surfaces.
\newblock {\em Journal of Computational Chemistry}, 29(10):1656--1666, July
  2008.

\bibitem{iori_golp_2009}
F.~Iori, R.~Di Felice, E.~Molinari, and S.~Corni.
\newblock {GolP}: {An} atomistic force-field to describe the interaction of
  proteins with {Au}(111) surfaces in water.
\newblock {\em Journal of Computational Chemistry}, 30(9):1465--1476, 2009.

\bibitem{pensado2011a}
A.~S. Pensado and A.~A.~H. P\'adua.
\newblock Solvation and stabilization of metallic nanoparticles in ionic
  liquids.
\newblock {\em Angew. Chem., Int. Ed.}, 50:8683--8687, 2011.

\bibitem{schmickler_interphase_1984}
Wolfgang Schmickler and Douglas Henderson.
\newblock The interphase between jellium and a hard sphere electrolyte. {A}
  model for the electric double layer.
\newblock {\em The Journal of Chemical Physics}, 80(7):3381--3386, April 1984.

\bibitem{shelley_modeling_1997}
J.~C. Shelley, G.~N. Patey, D.~R. Bérard, and G.~M. Torrie.
\newblock Modeling and structure of mercury-water interfaces.
\newblock {\em The Journal of Chemical Physics}, 107(6):2122--2141, August
  1997.

\bibitem{price_molecular_1995}
David~L. Price and J.~W. Halley.
\newblock Molecular dynamics, density functional theory of the
  metal--electrolyte interface.
\newblock {\em The Journal of Chemical Physics}, 102(16):6603--6612, April
  1995.

\bibitem{walbran1998a}
S.~Walbran, A.~Mazzolo, J.W. Halley, and D.L. Price.
\newblock Model for the electrostatic response of the copper -- water
  interface.
\newblock {\em J. Chem. Phys.}, 109(18):8076--8080, 1998.

\bibitem{finnis_interaction_1991}
M.~W. Finnis.
\newblock The interaction of a point charge with an aluminium (111) surface.
\newblock {\em Surface Science}, 241(1):61--72, January 1991.

\bibitem{finnis_interaction_1995}
M.~W. Finnis, R.~Kaschner, C.~Kruse, J.~Furthmuller, and M.~Scheffler.
\newblock The interaction of a point charge with a metal surface: theory and
  calculations for (111), (100) and (110) aluminium surfaces.
\newblock {\em Journal of Physics: Condensed Matter}, 7(10):2001--2019, March
  1995.
\newblock Publisher: IOP Publishing.

\bibitem{nalewajski1984a}
R.~F. Nalewajski.
\newblock Electrostatic effects in interactions between hard (soft) acids and
  bases.
\newblock {\em J. Am. Chem. Soc.}, 106:944--945, 1984.

\bibitem{mortier_electronegativity-equalization_1986}
Wilfried~J. Mortier, Swapan~K. Ghosh, and S.~Shankar.
\newblock Electronegativity-equalization method for the calculation of atomic
  charges in molecules.
\newblock {\em Journal of the American Chemical Society}, 108(15):4315--4320,
  July 1986.

\bibitem{rappe1991a}
A.~K. Rappe and W.~A. {Goddard III}.
\newblock Charge equilibration for molecular dynamics simulations.
\newblock {\em J. Phys. Chem.}, 95:3358--3363, 1991.

\bibitem{gingrich2010a}
T.~R. Gingrich and M.~Wilson.
\newblock On the ewald summation of gaussian charges for the simulation of
  metallic surfaces.
\newblock {\em Chem. Phys. Lett.}, 500(1--3):178--183, 2010.

\bibitem{coretti_mass-zero_2020}
A.~Coretti, L.~Scalfi, C.~Bacon, B.~Rotenberg, R.~Vuilleumier, G.~Ciccotti,
  M.~Salanne, and S.~Bonella.
\newblock Mass-zero constrained molecular dynamics for electrode charges in
  simulations of electrochemical systems.
\newblock {\em The Journal of Chemical Physics}, 152(19):194701, May 2020.
\newblock Publisher: American Institute of Physics.

\bibitem{streitz_electrostatic_1994}
F.~H. Streitz and J.~W. Mintmire.
\newblock Electrostatic potentials for metal-oxide surfaces and interfaces.
\newblock {\em Physical Review B}, 50(16):11996--12003, October 1994.
\newblock Publisher: American Physical Society.

\bibitem{onofrio2015b}
N.~Onofrio and A.~Strachan.
\newblock Voltage equilibration for reactive atomistic simulations of
  electrochemical processes.
\newblock {\em J. Chem. Phys.}, 143:054109, 2015.

\bibitem{onofrio2015a}
N.~Onofrio, D.~Guzman, and A.~Strachan.
\newblock Atomic origin of ultrafast resistance switching in nanoscale
  electrometallization cells.
\newblock {\em Nat. Mater.}, 14:440--446, 2015.

\bibitem{liang_applied_2018}
Tao Liang, Andrew~C. Antony, Sneha~A. Akhade, Michael~J. Janik, and Susan~B.
  Sinnott.
\newblock Applied {Potentials} in {Variable}-{Charge} {Reactive} {Force}
  {Fields} for {Electrochemical} {Systems}.
\newblock {\em The Journal of Physical Chemistry A}, 122(2):631--638, January
  2018.

\bibitem{nakano_chemical_2019}
Hiroshi Nakano and Hirofumi Sato.
\newblock A chemical potential equalization approach to constant potential
  polarizable electrodes for electrochemical-cell simulations.
\newblock {\em The Journal of Chemical Physics}, 151(16):164123, October 2019.

\bibitem{buraschi2020a}
M.~Buraschi, S.~Sansotta, and D.~Zahn.
\newblock Polarization effects in dynamic interfaces of platinum electrodes and
  ionic liquid phases: A molecular dynamics study.
\newblock {\em J. Phys. Chem. C}, 124:2002--2007, 2020.

\bibitem{nistor_dielectric_2009}
Razvan~A. Nistor and Martin~H. M\"user.
\newblock Dielectric properties of solids in the regular and split-charge
  equilibration formalisms.
\newblock {\em Physical Review B}, 79(10):104303, March 2009.

\bibitem{pastewka_charge-transfer_2011}
Lars Pastewka, Tommi~T. Järvi, Leonhard Mayrhofer, and Michael Moseler.
\newblock Charge-transfer model for carbonaceous electrodes in polar
  environments.
\newblock {\em Physical Review B}, 83(16):165418, April 2011.

\bibitem{salanne2015a}
M.~Salanne.
\newblock {Simulations of Room Temperature Ionic Liquids: from Polarizable to
  Coarse-Grained Force Fields}.
\newblock {\em Phys. Chem. Chem. Phys.}, 17:14270--14279, 2015.

\bibitem{merlet2013a}
C.~Merlet, M.~Salanne, B.~Rotenberg, and P.~A. Madden.
\newblock Influence of solvation on the structural and capacitive properties of
  electrical double layer capacitors.
\newblock {\em Electrochim. Acta}, 101:262--271, 2013.

\bibitem{tazi2010a}
S.~Tazi, M.~Salanne, C.~Simon, P.~Turq, M.~Pounds, and P.~A. Madden.
\newblock Potential-induced ordering transition of the adsorbed layer at the
  ionic liquid / electrified metal interface.
\newblock {\em J. Phys. Chem. B}, 114:8453--8459, 2010.

\bibitem{bedrov2019a}
D.~Bedrov, J.-P. Piquemal, O.~Borodin, A.~D. {MacKerell, Jr.}, B.~Roux, and
  C.~Schr\"oder.
\newblock Molecular dynamics simulations of ionic liquids and electrolytes
  using polarizable force fields.
\newblock {\em Chem. Rev.}, 119:7940--7995, 2019.

\bibitem{park_interference_2020}
Suehyun Park and Jesse~G. McDaniel.
\newblock Interference of electrical double layers: {Confinement} effects on
  structure, dynamics, and screening of ionic liquids.
\newblock {\em The Journal of Chemical Physics}, 152(7):074709, February 2020.

\bibitem{lebreton2020a}
G.~{Le Breton} and L.~Joly.
\newblock Molecular modeling of aqueous electrolytes at interfaces: Effects of
  long-range dispersion forces and of ionic charge rescaling.
\newblock {\em J. Chem. Phys.}, 152:241102, 2020.

\bibitem{russier_adsorption_1987}
V.~Russier, M.~L. Rosinberg, J.~P. Badiali, D.~Levesque, and J.~J. Weis.
\newblock Adsorption of polar molecules at a wall: {Monte} {Carlo} simulations
  and integral equations.
\newblock {\em The Journal of Chemical Physics}, 87(8):5012--5020, October
  1987.
\newblock Publisher: American Institute of Physics.

\bibitem{spohr_molecular_1986}
E.~Spohr and K.~Heinzinger.
\newblock Molecular dynamics simulation of a water/metal interface.
\newblock {\em Chemical Physics Letters}, 123(3):218--221, January 1986.

\bibitem{spohr_computer_1989}
E.~Spohr.
\newblock Computer simulation of the water/platinum interface.
\newblock {\em The Journal of Physical Chemistry}, 93(16):6171--6180, August
  1989.

\bibitem{alhamdani2017a}
Yasmine~S. Al-Hamdani, Dario Alf\`e, and Angelos Michaelides.
\newblock How strongly do hydrogen and water molecules stick to carbon
  nanomaterials?
\newblock {\em The Journal of Chemical Physics}, 146(9):094701, 2017.

\bibitem{heinz2008a}
H.~Heinz, R.~A. Vaia, B.~L. Farmer, and R.~R. Naik.
\newblock Accurate simulation of surfaces and interfaces of face--centered
  cubic metals using 12--6 and 9--6 lennard-jones potentials.
\newblock {\em J. Phys. Chem. C}, 112:17281--17290, 2008.

\bibitem{clabaut2020a}
P.~Clabaut, P.~{Fleurat-Lessard}, C.~Michel, and S.~N. Steinmann.
\newblock Ten facets, one force field: The {GAL19} force field for water--noble
  metal interfaces.
\newblock {\em J. Chem. Theory Comput.}, 16:4565--4578, 2020.

\bibitem{Marin-Lafleche2020}
Abel Marin-Lafl\`eche, Matthieu Haefele, Laura Scalfi, Alessandro Coretti,
  Thomas Dufils, Guillaume Jeanmairet, Stewart Reed, Alessandra Serva, Roxanne
  Berthin, Camille Bacon, Sara Bonella, Benjamin Rotenberg, Paul~Anthony
  Madden, and Mathieu Salanne.
\newblock {{MetalWalls}: A Classical Molecular Dynamics Software Dedicated to
  the Simulation of Electrochemical Systems (doi:
  10.26434/chemrxiv.12389777.v1)}.
\newblock june 2020.

\bibitem{plimpton1995a}
Steve Plimpton.
\newblock {Fast Parallel Algorithms for Short-Range Molecular Dynamics}.
\newblock {\em J. Comput. Phys.}, 117:1--19, 1995.

\bibitem{eastman_openmm_2017}
Peter Eastman, Jason Swails, John~D. Chodera, Robert~T. McGibbon, Yutong Zhao,
  Kyle~A. Beauchamp, Lee-Ping Wang, Andrew~C. Simmonett, Matthew~P. Harrigan,
  Chaya~D. Stern, Rafal~P. Wiewiora, Bernard~R. Brooks, and Vijay~S. Pande.
\newblock {OpenMM} 7: {Rapid} development of high performance algorithms for
  molecular dynamics.
\newblock {\em PLOS Computational Biology}, 13(7):e1005659, July 2017.
\newblock Publisher: Public Library of Science.

\bibitem{weik_espresso_2019}
Florian Weik, Rudolf Weeber, Kai Szuttor, Konrad Breitsprecher, Joost de~Graaf,
  Michael Kuron, Jonas Landsgesell, Henri Menke, David Sean, and Christian
  Holm.
\newblock {ESPResSo} 4.0, an extensible software package for simulating soft
  matter systems.
\newblock {\em The European Physical Journal Special Topics},
  227(14):1789--1816, March 2019.

\bibitem{AllCodes}
{Some open source classical molecular dynamics simulation codes allowing the
  simulation of electrodes}.
\newblock {Metalwalls: \url{https://gitlab.com/ampere2/metalwalls}; {LAMMPS}:
  \url{https://lammps.sandia.gov} for the distribution,
  \url{https://github.com/zhenxingwang/lammps-conp} for the constant-potential
  fix; {OpenMM}: \url{http://openmm.org} for the distribution,
  \url{https://github.com/scychon/openmm_constV} for the constant-potential
  plug-in; {ESPRESSO}: \url{espressomd.org}.}, note = {Accessed: 2020-08-20}.

\bibitem{yeh1999a}
I.~C. Yeh and M.~L. Berkowitz.
\newblock Ewald summation for systems with slab geometry.
\newblock {\em J. Chem. Phys.}, 111:3155--3162, 1999.

\bibitem{dufils_simulating_2019}
Thomas Dufils, Guillaume Jeanmairet, Benjamin Rotenberg, Michiel Sprik, and
  Mathieu Salanne.
\newblock Simulating {Electrochemical} {Systems} by {Combining} the {Finite}
  {Field} {Method} with a {Constant} {Potential} {Electrode}.
\newblock {\em Physical Review Letters}, 123(19):195501, November 2019.
\newblock Publisher: American Physical Society.

\bibitem{stengel2009b}
M.~Stengel, N.~A. Spaldin, and D.~Vanderbilt.
\newblock Electric displacement as the fundamental variable in
  electronic-structure calculations.
\newblock {\em Nature Phys.}, 5:304--308, 2009.

\bibitem{zhang2016f}
C.~Zhang and M.~Sprik.
\newblock Finite field methods for the supercell modeling of charged
  insulator/electrolyte interfaces.
\newblock {\em Phys. Rev. B}, 94:245309, 2016.

\bibitem{sayer2017a}
T.~Sayer, C.~Zhang, and M.~Sprik.
\newblock Charge compensation at the interface between the polar {NaCl}(111)
  surface and a {NaCl} aqueous solution.
\newblock {\em J. Chem. Phys.}, 147:104702, 2017.

\bibitem{bonnet_first-principles_2012}
Nicéphore Bonnet, Tetsuya Morishita, Osamu Sugino, and Minoru Otani.
\newblock First-{Principles} {Molecular} {Dynamics} at a {Constant} {Electrode}
  {Potential}.
\newblock {\em Physical Review Letters}, 109(26):266101, December 2012.
\newblock Publisher: American Physical Society.

\bibitem{punnathanam_gibbs-ensemble_2014}
Sudeep~N. Punnathanam.
\newblock A {Gibbs}-ensemble based technique for {Monte} {Carlo} simulation of
  electric double layer capacitors ({EDLC}) at constant voltage.
\newblock {\em The Journal of Chemical Physics}, 140(17):174110, May 2014.
\newblock Publisher: American Institute of Physics.

\bibitem{stenberg_grand_2020}
Samuel Stenberg, Björn Stenqvist, Cliff Woodward, and Jan Forsman.
\newblock Grand canonical simulations of ions between charged conducting
  surfaces using exact {3D} {Ewald} summations.
\newblock {\em Physical Chemistry Chemical Physics}, 22(24):13659--13665, June
  2020.
\newblock Publisher: The Royal Society of Chemistry.

\bibitem{haskins2016a}
J.~B. Haskins and J.~W. Lawson.
\newblock Evaluation of molecular dynamics simulation methods for ionic liquid
  electric double layers.
\newblock {\em J. Chem. Phys.}, 144:134701, 2016.

\bibitem{scalfi2020a}
L.~Scalfi, D.~T. Limmer, A.~Coretti, S.~Bonella, P.~A. Madden, M.~Salanne, and
  B.~Rotenberg.
\newblock Charge fluctuations from molecular simulations in the
  constant-potential ensemble.
\newblock {\em Phys. Chem. Chem. Phys.}, 22:10480--10489, 2020.

\bibitem{limmer2013a}
D.~T. Limmer, C.~Merlet, M.~Salanne, D.~Chandler, P.~A. Madden, R.~{van Roij},
  and B.~Rotenberg.
\newblock Charge fluctuations in nanoscale capacitors.
\newblock {\em Phys. Rev. Lett.}, 111:106102, 2013.

\bibitem{takahashi_polarizable_2020}
Ken Takahashi, Hiroshi Nakano, and Hirofumi Sato.
\newblock A polarizable molecular dynamics method for electrode--electrolyte
  interfacial electron transfer under the constant
  chemical-potential-difference condition on the electrode electrons.
\newblock {\em The Journal of Chemical Physics}, 153(5):054126, August 2020.
\newblock Publisher: American Institute of Physics.

\bibitem{patel_fluctuations_2010}
Amish~J. Patel, Patrick Varilly, and David Chandler.
\newblock Fluctuations of {Water} near {Extended} {Hydrophobic} and
  {Hydrophilic} {Surfaces}.
\newblock {\em The Journal of Physical Chemistry B}, 114(4):1632--1637,
  February 2010.

\bibitem{kattirtzi2017a}
John~A. Kattirtzi, David~T. Limmer, and Adam~P. Willard.
\newblock Microscopic dynamics of charge separation at the aqueous
  electrochemical interface.
\newblock {\em Proceedings of the National Academy of Sciences},
  114(51):13374--13379, December 2017.

\bibitem{willard2009a}
A.~P. Willard, S.~K. Reed, P.~A. Madden, and D.~Chandler.
\newblock Water at an electrochemical interface - a simulation study.
\newblock {\em Faraday Discuss.}, 141:423--441, 2009.

\bibitem{bonthuis2012a}
Douwe~Jan Bonthuis, Stephan Gekle, and Roland~R. Netz.
\newblock Profile of the {Static} {Permittivity} {Tensor} of {Water} at
  {Interfaces}: {Consequences} for {Capacitance}, {Hydration} {Interaction} and
  {Ion} {Adsorption}.
\newblock {\em Langmuir}, 28(20):7679--7694, 2012.

\bibitem{jeanmairet2019b}
Guillaume Jeanmairet, Benjamin Rotenberg, Daniel Borgis, and Mathieu Salanne.
\newblock Study of a water-graphene capacitor with molecular density functional
  theory.
\newblock {\em The Journal of Chemical Physics}, 151(12):124111, September
  2019.

\bibitem{li2018b}
Zhujie Li, Guillaume Jeanmairet, Trinidad Mendez-Morales, Benjamin Rotenberg,
  and Mathieu Salanne.
\newblock Capacitive {Performance} of {Water}-in-{Salt} {Electrolytes} in
  {Supercapacitors}: {A} {Simulation} {Study}.
\newblock {\em The Journal of Physical Chemistry C}, 122(42):23917--23924,
  2018.

\bibitem{suo2013a}
L.~Suo, Y.-S. Hu, H.~Li, M.~Armand, and L.~Chen.
\newblock {A New Class of Solvent-in-Salt Electrolyte for High-Energy
  Rechargeable Metallic Lithium Batteries}.
\newblock {\em Nat. Commun.}, 4:1481, 2013.

\bibitem{wang2016j}
J.~Wang, Y.~Yamada, K.~Sodeyama, C.~H. Chiang, Y.~Tateyama, and A.~Yamada.
\newblock {Superconcentrated Electrolytes for a High-voltage Lithium-ion
  Battery}.
\newblock {\em Nat. Commun.}, 7:12032, 2016.

\bibitem{rotenberg2015a}
B.~Rotenberg and M.~Salanne.
\newblock Structural transitions at ionic liquid interfaces.
\newblock {\em J. Phys. Chem. Lett.}, 6:4978--4985, 2015.

\bibitem{uralcan2016a}
B.~Uralcan, I.~A. Aksay, P.~G. Debenedetti, and D.~T. Limmer.
\newblock Concentration fluctuations and capacitive response in dense ionic
  solutions.
\newblock {\em J. Phys. Chem. Lett.}, asap, 2016.

\bibitem{pean2014a}
C.~P\'ean, C.~Merlet, B.~Rotenberg, P.~A. Madden, P.-L. Taberna, B.~Daffos,
  M.~Salanne, and P.~Simon.
\newblock On the dynamics of charging in nanoporous carbon-based
  supercapacitors.
\newblock {\em ACS Nano}, 8:1576--1583, 2014.

\bibitem{pean2015b}
C.~Pean, B.~Daffos, B.~Rotenberg, P.~Levitz, M.~Haefele, P.-L. Taberna,
  P.~Simon, and M.~Salanne.
\newblock Confinement, desolvation, and electrosorption effects on the
  diffusion of ions in nanoporous carbon electrodes.
\newblock {\em J. Am. Chem. Soc.}, 137:12627, 2015.

\bibitem{pean2016b}
C.~Pean, B.~Rotenberg, P.~Simon, and M.~Salanne.
\newblock Multi-scale modelling of supercapacitors: From molecular simulations
  to a transmission line model.
\newblock {\em J. Power Sources}, 326:680--685, 2016.

\bibitem{breitsprecher2018a}
Konrad Breitsprecher, Christian Holm, and Svyatoslav Kondrat.
\newblock Charge {Me} {Slowly}, {I} {Am} in a {Hurry}: {Optimizing}
  {Charge}/{Discharge} {Cycles} in {Nanoporous} {Supercapacitors}.
\newblock {\em ACS Nano}, 12(10):9733--9741, October 2018.

\bibitem{limmer2013b}
D.~T. Limmer, A.~P. Willard, P.~Madden, and D.~Chandler.
\newblock Hydration of metal surfaces can be dynamically heterogeneous and
  hydrophobic.
\newblock {\em Proc. Natl. Acad. Sci. U.S.A.}, 110:4200--4205, 2013.

\bibitem{mullen2014a}
Ryan~Gotchy Mullen, Joan-Emma Shea, and Baron Peters.
\newblock Transmission {Coefficients}, {Committors}, and {Solvent}
  {Coordinates} in {Ion}-{Pair} {Dissociation}.
\newblock {\em Journal of Chemical Theory and Computation}, 10(2):659--667,
  2014.

\bibitem{willard2013a}
A.~P. Willard, D.~T. Limmer, P.~A. Madden, and D.~Chandler.
\newblock Characterizing heterogeneous dynamics at hydrated electrode surfaces.
\newblock {\em J. Chem. Phys.}, 138:184702, 2013.

\bibitem{limmer2015b}
David~T. Limmer, Adam~P. Willard, Paul~A. Madden, and David Chandler.
\newblock Water {Exchange} at a {Hydrated} {Platinum} {Electrode} is {Rare} and
  {Collective}.
\newblock {\em The Journal of Physical Chemistry C}, 119(42):24016--24024,
  2015.

\bibitem{limmer2015c}
David~T. Limmer and Adam~P. Willard.
\newblock Nanoscale heterogeneity at the aqueous electrolyte--electrode
  interface.
\newblock {\em Chemical Physics Letters}, 620:144--150, January 2015.

\bibitem{zhang2020a}
Yiwei Zhang, Guillaume Stirnemann, James~T. Hynes, and Damien Laage.
\newblock Water dynamics at electrified graphene interfaces: a jump model
  perspective.
\newblock {\em Physical Chemistry Chemical Physics}, 22(19):10581--10591, 2020.
\newblock Publisher: The Royal Society of Chemistry.

\bibitem{choudhuri2016a}
J.~R. Choudhuri, D.~Vanzo, P.~A. Madden, M.~Salanne, D.~Bratko, and A.~Luzar.
\newblock Dynamic response in nanoelectrowetting on a dielectric.
\newblock {\em ACS Nano}, 10:8536--8544, 2016.

\bibitem{krupenkin2003a}
T.~Krupenkin, S.~Yang, and P.~Mach.
\newblock Tunable liquid microlens.
\newblock {\em Appl. Phys. Lett.}, 82:316, 2003.

\bibitem{marcus1956a}
R.~A. Marcus.
\newblock On the theory of oxidation-reduction reactions involving electron
  transfer. i.
\newblock {\em J. Chem. Phys.}, 24:966--978, 1956.

\bibitem{warshel1982a}
A.~Warshel.
\newblock Dynamics of reactions in polar solvents. semiclassical trajectory
  studies of electron-transfer and proton-transfer reactions.
\newblock {\em J. Phys. Chem.}, 86:2218--2224, 1982.

\bibitem{reed2008a}
S.~K. Reed, P.~A. Madden, and A.~Papadopoulos.
\newblock Electrochemical charge transfer at a metallic electrode: a simulation
  study.
\newblock {\em J. Chem. Phys.}, 128(12):124701, 2008.

\bibitem{pounds2015a}
M.~A. Pounds, M.~Salanne, and P.~A. Madden.
\newblock Molecular aspects of the {Eu}$^{3+}$/{Eu}$^{2+}$ redox reaction at
  the interface between a molten salt and a metallic electrode.
\newblock {\em Mol. Phys.}, 113:2451--2462, 2015.

\bibitem{wilhelm_modeling_2011}
F.~Wilhelm, W.~Schmickler, R.~Nazmutdinov, and E.~Spohr.
\newblock Modeling proton transfer to charged silver electrodes.
\newblock {\em Electrochimica Acta}, 56(28):10632--10644, December 2011.

\bibitem{wiebe_water_2017}
Johannes Wiebe and Eckhard Spohr.
\newblock Water {Structure} and {Mechanisms} of {Proton} {Discharge} on
  {Platinum} {Electrodes}: {Empirical} {Valence} {Bond} {Molecular} {Dynamics}
  {Trajectory} {Studies}.
\newblock {\em Electrocatalysis}, 8(6):637--646, November 2017.

\bibitem{dwelle2019a}
K.~A. Dwelle and A.~P. Willard.
\newblock Constant potential, electrochemically active boundary conditions for
  electrochemical simulation.
\newblock {\em J. Phys. Chem. C}, 123:24095--24103, 2019.

\bibitem{rose_solvation_1991}
Daniel~A. Rose and Ilan Benjamin.
\newblock Solvation of {Na}$^+$ and {Cl}$^-$ at the water--platinum (100)
  interface.
\newblock {\em The Journal of Chemical Physics}, 95(9):6856--6865, November
  1991.

\bibitem{calhoun_electron_1996}
August Calhoun and Gregory~A. Voth.
\newblock Electron {Transfer} {Across} the {Electrode}/{Electrolyte}
  {Interface}: {Influence} of {Redox} {Ion} {Mobility} and {Counterions}.
\newblock {\em The Journal of Physical Chemistry}, 100(25):10746--10753,
  January 1996.

\bibitem{TFarXiv}
Thomas Dufils, Laura Scalfi, Benjamin rotenberg, and Mathieu Salanne.
\newblock A semiclassical thomas-fermi model to tune the metallicity of
  electrodes in molecular simulations, 2019.

\bibitem{schlaich2020arXiv}
Alexander Schlaich, Dongliang Jin, Lydéric Bocquet, and Benoit Coasne.
\newblock Wetting transition of ionic liquids at metal surfaces: A
  computational approach to electronic screening using a virtual thomas-fermi
  fluid, 2020.

\bibitem{comtet2017a}
J.~Comtet, A.~Nigu\`es, V.~Kaiser, B.~Coasne, L.~Bocquet, and A.~Siria.
\newblock Nanoscale capillary freezing of ionic liquids confined between
  metallic interfaces and the role of electronic screening.
\newblock {\em Nat. Mater.}, 16:634--639, 2017.

\bibitem{jeanmairet2019a}
G.~Jeanmairet, B.~Rotenberg, M.~Levesque, D.~Borgis, and M.~Salanne.
\newblock A molecular density functional theory approach to electron transfer
  reactions.
\newblock {\em Chem. Sci.}, 10:2130--2143, 2019.

\bibitem{fajardo2015a}
O.~Y. Fajardo, F.~Bresme, A.~A. Kornyshev, and M.~Urbakh.
\newblock Electrotunable lubricity with ionic liquid nanoscale films.
\newblock {\em Sci. Rep.}, 5:7698, 2015.

\end{thebibliography}
\end{document}